# Observation of robust zero-energy state and enhanced superconducting gap in a tri-layer heterostructure of MnTe/Bi$_2$Te$_3$/Fe(Te, Se)

Shuyue Ding[1†], Chen Chen[1†], Zhipeng Cao[2†], Di Wang[2†], Yongqiang Pan[3], Ran Tao[1], Dongming Zhao[1], Yining Hu[1], Tianxing Jiang[1], Yajun Yan[4], Zhixiang Shi[3], Xiangang Wan[2,5], Donglai Feng[4,5,6,7], Tong Zhang[1,5,6*]

[1] Department of Physics, State Key Laboratory of Surface Physics and Advanced Material Laboratory, Fudan University, Shanghai 200438, China.
[2] National Laboratory of Solid State Microstructures and School of Physics, Nanjing University, Nanjing 210093, China.
[3] School of Physics and Key Laboratory of MEMS of the Ministry of Education, Southeast University, Nanjing 211189, China.
[4] Department of Physics, University of Science and Technology of China, Hefei 230026, China.
[5] Collaborative Innovation Center of Advanced Microstructures, Nanjing 210093, China.
[6] Shanghai Research Center for Quantum Sciences, Shanghai 201315, China.
[7] National Synchrotron Radiation Laboratory, University of Science and Technology of China, Hefei 230026, China.

† These authors contributed equally to this work
* Email: tzhang18@fudan.edu.cn

The interface between magnetic material and superconductors has long been predicted to host unconventional superconductivity, such as spin-triplet pairing and topological nontrivial pairing state, particularly when spin-orbital coupling (SOC) is incorporated. To identify these novel pairing states, fabricating homogenous heterostructures which contain such various properties are preferred, but often challenging. Here we synthesized a tri-layer type van-der Waals heterostructure of MnTe/Bi$_2$Te$_3$/Fe(Te, Se), which combined $s$-wave superconductivity, thickness dependent magnetism and strong SOC. Via low-temperature scanning tunneling microscopy (STM), we observed robust zero-energy states with notably nontrivial properties and an enhanced superconducting gap size on single unit-cell (UC) MnTe surface. In contrast, no zero-energy state was observed on 2UC MnTe. First-principle calculations further suggest the 1UC MnTe has large interfacial Dzyaloshinskii-Moriya interaction (DMI) and a frustrated AFM state, which could promote non-collinear spin textures. It thus provides a promising platform for exploring topological nontrivial superconductivity.

**Introduction:**

When materials with different quantum orders meet at the interface, new physical phenomena may emerge and bring potential applications. A well-known example is the ferromagnet - ($s$-wave) superconductor interface. Intensive studies have shown that while the spin-singlet pairing potential decays fast in ferromagnet, an odd-frequency spin-triplet pairing state can be induced (1-2). A spin rotation layer or spin orbital coupling (SOC) can further stabilize triplet Cooper pair and generate spin-polarized supercurrent (3-6), which is of particular interest in spintronic applications. Recently, there are increased research on magnet-superconductor interface with strong SOC or non-collinear spin textures, as these systems may give rise to topological superconductivity (TSC) (7-19). The chiral magnetic atom chain grown on strong SOC superconductors was considered as its 1D case (7,8), which is in analogy to the semiconductor nanowire-superconductor hybrids under Zeeman field (20,21). In the 2D case, magnetic moments arranged on superconductor can form Shiba lattice (9,10). Various types of chiral spin textures are predicted to induce effective *p*-wave pairing, and Majorana boundary mode will emerge when the system enters topological phase. The merits of these potential TSC platforms are that they don't require external field and the spin textures could be controlled electronically (22,23). Besides, theoretical works also predicted that if a 3D topological insulator (TI) layer is inserted in between ferromagnet and superconductor layers, 2D TSC/chiral Majorana edge mode can be induced due to different gap opening in the top/bottom surface states (24,25).

To experimentally identify these unconventional pairing states, fabrication of high quality magnet-superconductor heterostructures with proper interface is highly desired, since unconventional pairing is often sensitive to disorder or inhomogeneity, and local probe study such as STM favors atomically flat surface/interface. Several recent STM studies (26-29) have demonstrated the advantage of using layered magnetic material (like transition metal chalcogenides) to make such heterostructures, as they can be grown epitaxially and persist magnetism down to single-layer limit. Moreover, it is also preferred to introduce strong SOC material or topological band structure (such as a TI) into the heterostructure, as they can further promote the TSC phase. Particularly, it was shown that if combining strong SOC and inversion asymmetry (which always happens at interface) in a magnetic system, large Dzyaloshinskii-Moriya interactions (DMI) can be induced and give rise to non-collinear spin textures (30-32). Transport studies on ferromagnet-TI interfaces have indeed observed signatures of skyrmion-like spin structure (33-35). Therefore, it would be very promising to fabricate a magnet-superconductor heterostructure incorporated with strong SOC materials for exploring potential unconventional superconductivity. However so far such kind of heterostructure is rarely reported.

In this work, we successfully synthesized a tri-layer type heterostructure of MnTe/$Bi_2Te_3$/Fe(Te, Se) which possesses atomically sharp interface. It combines the magnetism of MnTe, strong SOC from $Bi_2Te_3$ and superconductivity from Fe(Te, Se). Via low-temperature STM, we observed zero-energy conductance peaks (ZECP) on many regions of one unit cell (UC) thick MnTe surface, and an enlarged superconducting gap with comparing to that of Fe(Te, Se). The ZECPs show several distinguished properties: they are not correlated to the observed surface defects, robust under out-of-plane field and display saturated conductance at low tunneling barriers. Meanwhile, no in-gap state was observed on 2 UC MnTe surface. Our first-principle calculations suggest that 1UC MnTe with the $Bi_2Te_3$ host large interfacial DMI and a frustrated AFM state, which can promote non-collinear spin textures. Thus the emergence of ZECPs and their nontrivial features may originate from topological pairing state.

**Results:**

**Growth and STM characterization of the heterostructure**

MnTe is a layered antiferromagnetic (AFM) insulator with a NiAs-type hexagonal structure (36,37) (which has an in-plane ferromagnetic order and AFM coupling between each plane); while bulk $Bi_2Te_3$ is a 3D TI whose unit-cell has a "quintuple layer" structure along c-axis (38). We fabricated the $MnTe/Bi_2Te_3/Fe(Te, Se)$ heterostructure via ultrahigh vacuum molecular beam epitaxy (UHV-MBE). 1 UC $Bi_2Te_3$ was firstly grown on as-cleaved Fe(Te, Se) surface, followed by the growth of 1~2 UC MnTe film and post annealing. The sample was then transferred to a low-temperature STM with calibrated resolution (Fig. S1). Detailed experimental procedures are described in "Materials and Methods" section. The 1UC $Bi_2Te_3$ here not only provides strong SOC to the interface, but also acts as a necessary buffer layer for MnTe growth, as its in-plane lattice constant (0.437 nm) is close to MnTe (0.419 nm) and has a Te-terminated surface. Moreover, previous studies have shown that $Bi_2Te_3$ has good superconducting proximity effect when grown on Fe(Te, Se) (39,40).

Fig. 1A shows a large-scale STM image of as-grown $MnTe/Bi_2Te_3/Fe(Te, Se)$ heterostructure. The surface contains various types of atomically flat terraces, including 1UC $Bi_2Te_3$, 1UC / 2UC MnTe on 1UC $Bi_2Te_3$, and bare Fe(Te, Se) substrate. They can be identified through the step height (reflected by different colors in Fig. 1A) and the in-plane lattice structure. Fig. 1B shows a line profile taken across several terraces in Fig. 1A. The 1UC $Bi_2Te_3$ has its characteristic height of ~1.0 nm and 1UC MnTe has a height of 0.32 nm, which corresponds to the single UC thickness in their bulk form (37,38). Figs. 1C, 1D are zoomed-in images of 1UC $Bi_2Te_3$ and 1UC MnTe surfaces while Figs. 1E, 1F show their atomic lattice, respectively. Both of the two surfaces have a hexagonal lattice with a constant of 0.43 nm and are oriented in the same direction, which indicates MnTe is epitaxially grown on $Bi_2Te_3$ (sketched in Fig. 1B). Meanwhile, the exposed Fe(Te, Se) remains intact with a square lattice (Fig. 1G), indicative of a sharp interface. We note this $MnTe/Bi_2Te_3$ epitaxial structure is different from the $MnBi_2Te_4$ film reported by ref. 41, which is also grown by MBE but with different growth/annealing conditions. Actually a similar epitaxial structure was reported in the growth of MnSe on $Bi_2Se_3$ surface (42). As seen in Figs. 1C, 1D, there are certain amount of point defects on both $Bi_2Te_3$ and MnTe surfaces, which locate at the hollow site of Te lattice, as indicated in Figs. 1E, 1F. They are thus likely the substitutional defects of Bi or Mn (43). Nonetheless, these defects do not affect the superconductivity in either MnTe or $Bi_2Te_3$ surface noticeably, as shown in Fig. S2 and further discussion below.

Having identified the interface structure, the electronic states are studied by tunneling spectroscopy. Fig. 1H shows large energy scale (±1 eV) dI/dV spectra taken on different surfaces. On Fe(Te, Se) substrate and 1 UC $Bi_2Te_3$, the line shapes of the spectra are similar to previous studies (39,40). For 1 UC MnTe on $Bi_2Te_3/Fe(Te, Se)$, the dI/dV displays a large band gap about 500 meV wide, which is consistent with that bulk MnTe is an insulator. However, the $E_F$ of 1 UC MnTe locates close to upper edge of the gap. This is likely due to electron doping effect induced by interfacial charge transfer. Figs. 2A,2B show the quasiparticle interference (QPI) measurement on 1UC MnTe which further revealed a shallow electron pocket crossed $E_F$, with a band bottom at -60(±10) meV (see Fig. S3 for additional QPI data). This indicates the electrons in 1UC MnTe will participate in superconductivity (QPI patterns are also observed on 1UC $Bi_2Te_3$, but displays a different dispersion, see Fig. S4). As for the 2 UC MnTe, there is still a flat insulating gap but the $E_F$ is at the gap center, indicating a weakened doping effect.

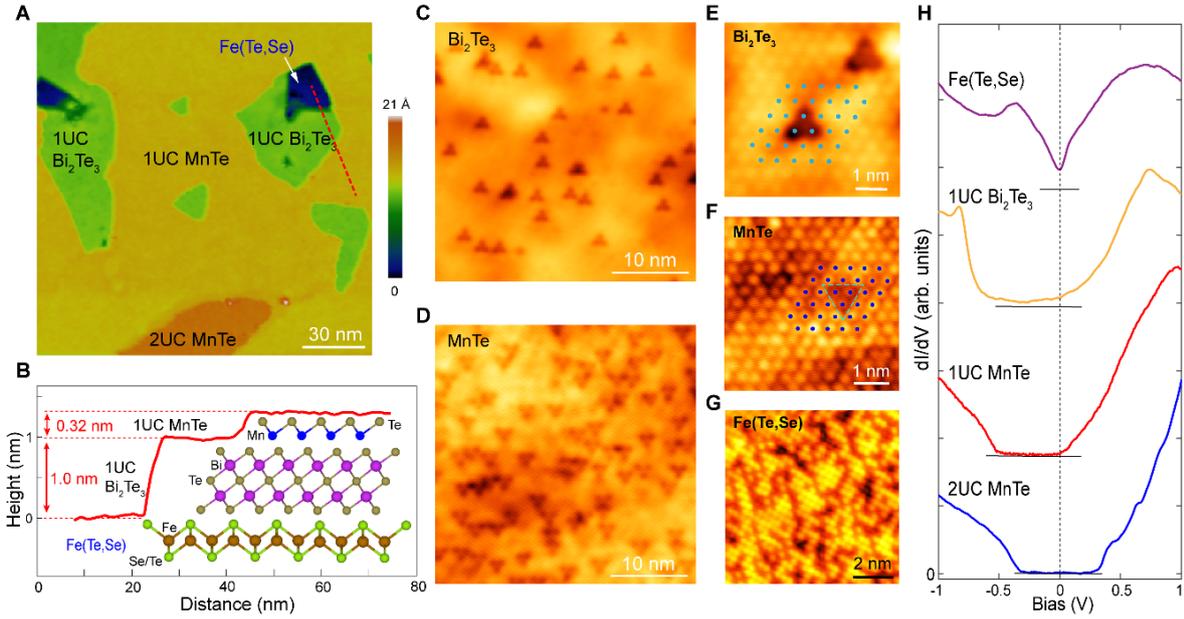

**Fig. 1. The surface of MnTe/$Bi_2Te_3$/Fe(Te, Se) heterostructure.** (**A**) Typical topographic image of the heterostructure (160×160 nm$^2$, $V_b$ = 1V, $I$ = 20pA). Different surface termination layers can be identified by different false colors. (**B**) A line profile taken along the dashed line in panel (A). The lattice structure of MnTe, $Bi_2Te_3$ layers and Fe(Te, Se) substrate are illustrated. (**C, D**) Topographic image of the 1UC $Bi_2Te_3$ ($V_b$ = 1V, $I$ =100 pA) and 1UC MnTe/$Bi_2Te_3$ surface ($V_b$ =1 V, $I$ =20 pA), respectively. (**E, F, G**) Atomically resolved images of MnTe, $Bi_2Te_3$ and exposed Fe(Te, Se) surface, respectively. The surface Te lattice is indicated on E and F. (**H**) Typical large scale dI/dV spectra taken on different surface terraces (setpoint: $V_b$ = 1V, $I$ = 100pA).

The low energy dI/dV spectra which reflect superconducting state are then measured at T = 0.4K on different surfaces, as summarized in Fig. 2C. The superconducting spectrum of Fe(Te, Se) substrate is fully gapped with multiple pairs of coherence peaks at ±1.5 meV, ±1.9 meV and ±2.5 meV, respectively. We note similar gaps were also reported in previous STM study and were assigned to different bands of Fe(Te, Se) (44). The gap spectrum of 1UC $Bi_2Te_3$ share similarities with Fe(Te, Se): its major coherence peaks are at ±1.9 meV while kink features at ±1.5 meV and ±2.5 meV can also be observed. This indicates a good superconducting proximity effect in $Bi_2Te_3$. However for 1UC MnTe, the spectrum displays rather unconventional behavior. Firstly, in about 50% regions a large "U" shaped gap with a size of 2.9 meV (defined by the lowest kink feature at gap edge) is observed, as the one shown in Fig. 2C (red curve). Fig. 2D displays how the gap evolves at a boundary between 1UC $Bi_2Te_3$ and 1UC MnTe, in which one can clearly see the gap enhancement in 1UC MnTe region. Fig. 2E shows the temperature dependence of the 1UC MnTe gap. It almost disappeared above 13.6K which cannot be ascribed to the thermo broadening only (by comparing with numerically broadened spectra in Fig. 2E). This indicates the gap of MnTe is still a proximity gap induced by Fe(Te, Se) (similar temperature dependence was also observed for the gap of 1UC $Bi_2Te_3$, see Fig. S5).

Moreover, we have also observed vortex state in the gapped region of 1UC MnTe (Fig. S6), which further confirms its superconducting nature. The enlarged gap size in 1UC MnTe is quite unexpected as one usually observes reduced gap in magnet/superconductor interfaces. Besides the enhanced gap, in other regions of 1UC MnTe we observed multiple in-gap states and zero-energy modes, which will be discussed in detail below. As for the 2UC MnTe (green curve in Fig. 2C), in contrast to 1UC MnTe it shows a uniform gap similar to $Bi_2Te_3$, with pronounced coherence peaks locate at ±1.9 meV.

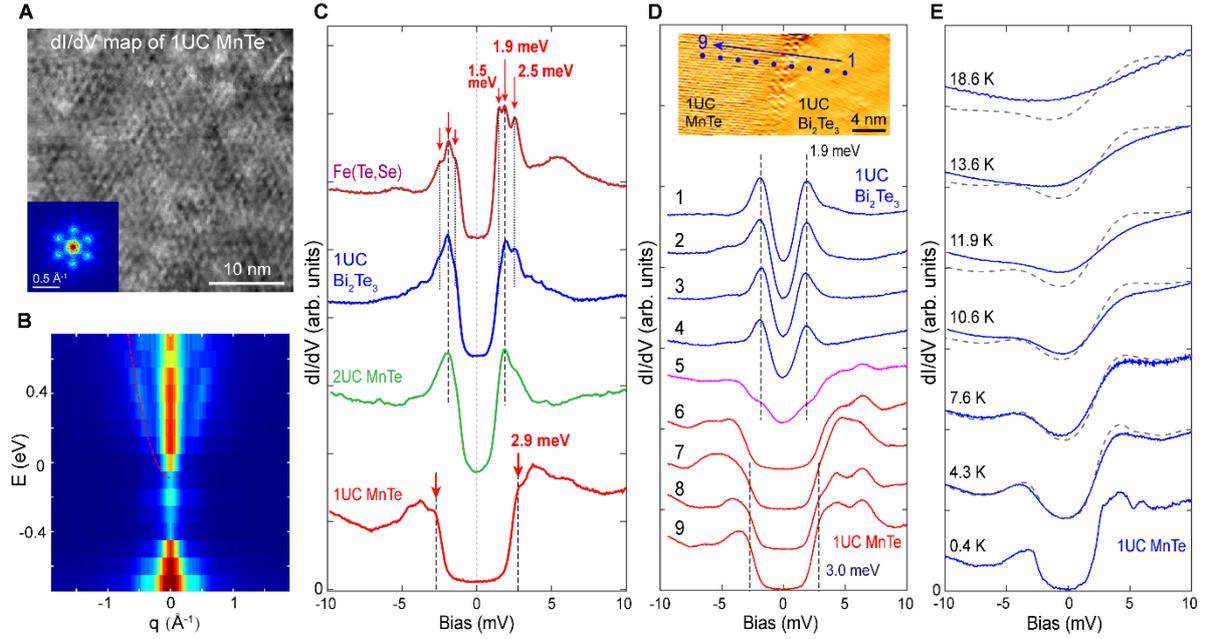

**Fig. 2. The tunneling spectra and QPI of MnTe/Bi$_2$Te$_3$/Fe(Te, Se) heterostructure.** (**A**) dI/dV map and its FFT image taken on 1UC MnTe surface ($V_b$ = 0.4 V, $I$ = 100pA). (**B**) Summary of the FFT line profiles taken at various energies, which shows an electron-like dispersion with a band bottom at $E_b \approx$ -60 meV. (**C**) Typical superconducting gap spectra taken on different surface layers. The positions of the multiple coherence peaks are marked by arrows and tracked by dashed lines. (**D**) A series of spectra taken across a boundary between 1UC MnTe and 1UC Bi$_2$Te$_3$ (Inset: topographic image of the boundary, the positions where the spectra were taken are marked) (**E**) Temperature dependence of the proximity superconducting gap on 1UC MnTe. The dashed curves are convolutions of the spectrum taken at T=0.4K with Fermi-Dirac distribution function at corresponding temperatures.

**In-gap states and zero-energy modes in 1UC MnTe/Bi$_2$Te$_3$/Fe(Te, Se)**

To further investigate the spatial distributions of superconducting states, Figs. 3A, 3B show a surface topography and its zero-bias dI/dV map (taken at B= 0T), respectively. This area contains different terraces including 1UC MnTe, 1UC Bi$_2$Te$_3$ and Fe(Te, Se) substrate. The zero-bias conductance (ZBC) in the 1UC Bi$_2$Te$_3$ and Fe(Te, Se) regions are all near zero, indicating a spatially uniform superconducting gap. However, on the 1UC MnTe surfaces, there are many separated bright regions with finite ZBC. These small regions have a typical scale of 4 to 10 nm and we refer them as quasiparticle "puddles" below. Figs. 3C, 3D show two series of dI/dV spectra taken across two puddles indicated in Fig. 3B. A clear ZECP together with multiple discrete low-energy states are seen. Their intensities decay as leaving the puddle but the energies do not change obviously. Figs. 3E-3N show the typical spectra taken in other 10 different puddles (spots 1-10 in Fig. 3B). In-gap bound states and ZECPs are commonly observed. Meanwhile, the spectra taken outside of puddles show clean "U" shaped gap with a size of ≈2.9 meV (spots 11-13, Fig. 3O). We have repeatedly observed the quasi-particle puddles with a ZECP in different 1UC MnTe terraces (an additional data set is shown in Fig. S7). By fitting the dI/dV spectra with Gaussian peaks, we summarized the appearance probability of the in-gap states at different energies in Fig. 3P, the ZECP has a probability over 80%. The peak width (FWHM) of the ZECP is in the range of 0.35-0.55 meV, which is close to the energy resolution of our system.

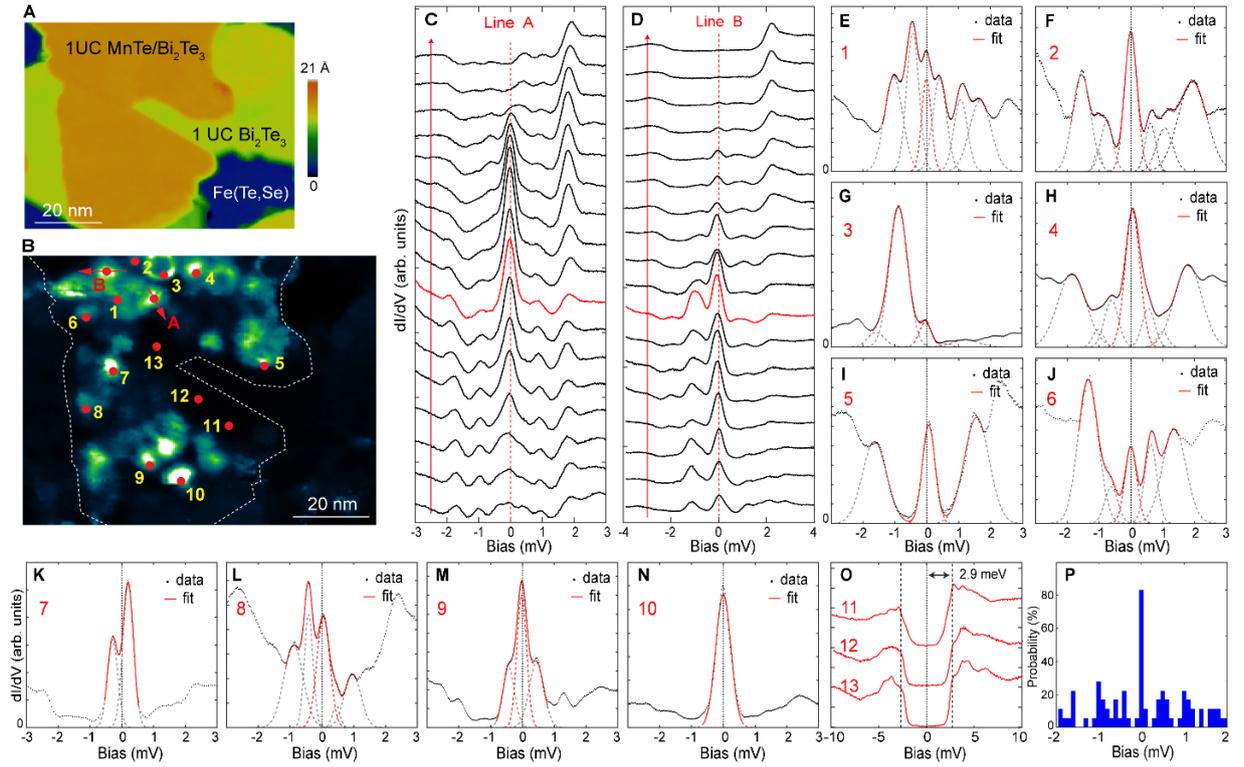

**Fig. 3. ZECP and in-gap states in 1UC MnTe/Bi$_2$Te$_3$/Fe(Te, Se).** (**A**) STM image of the heterostructure. (80×60 nm$^2$). (**B**) Zero-bias dI/dV map of the area in panel (A) at B = 0T (setpoint: $V_b$ =10 mV, $I$ =100 pA). The boundary of 1UC MnTe terrace is tracked by white curve. (**C, D**) dI/dV spectra taken along the red arrows *A* and *B* in panel (B), respectively (setpoint: $V_b$ = 3 mV, $I$ =200 pA). (**E-O**) dI/dV spectra measured on numbered spots in panel (B). The red solid curves in each panel are Gaussian peak fitting (dashed lines are individual peaks). (**P**) Statistics of the appearance probability of the in-gap states at different energies.

The quasiparticle puddles appear to distribute randomly. We have checked that the ZECP/in-gap state are not directly induced by observed defects in 1UC MnTe (see Figs. S2), and the distribution of quasiparticle puddles are also not correlated to surface defects or unevenness (see section S7 and Fig. S8). This indicates these in-gap states are unlikely the Yu-Shiba-Rusinov (YSR) state induced by impurities in MnTe. We also checked that the defects in exposed 1UC $Bi_2Te_3$ do not induce YSR state either (Fig. S2), which excludes the in-gap states in 1UC MnTe are induced by the defects of underlying $Bi_2Te_3$ layer.

To explore the origin of ZECP and the in-gap states, we further measured their response to external magnetic field, elevated temperature and lowered tunneling barrier. Fig. 4A shows the typical spectra measured under out-of-plane field. The ZECP persists and does not show splitting behavior up to B=10T (detailed analysis show that the ZECP is only slightly broadened by ~0.14 meV at B=10T, but significant shift or broadening is observed for the non-zero energy state, see Section S8 and Fig. S9). Meanwhile, the shape and spatial distribution of the quasiparticle puddles do not show obvious change either (Fig. S10). Such robustness against out-of-plane B field for ZECP also disfavors YSR-like states, which should split with a Zeeman energy (~1.1meV for B=10T). Fig. 4B shows a temperature dependence of the in-gap states spectra. All the in-gap states are broadened quickly at elevated temperatures and disappear near the $T_C$ of Fe(Te, Se), which indicates they are still quasiparticle states related to superconductivity.

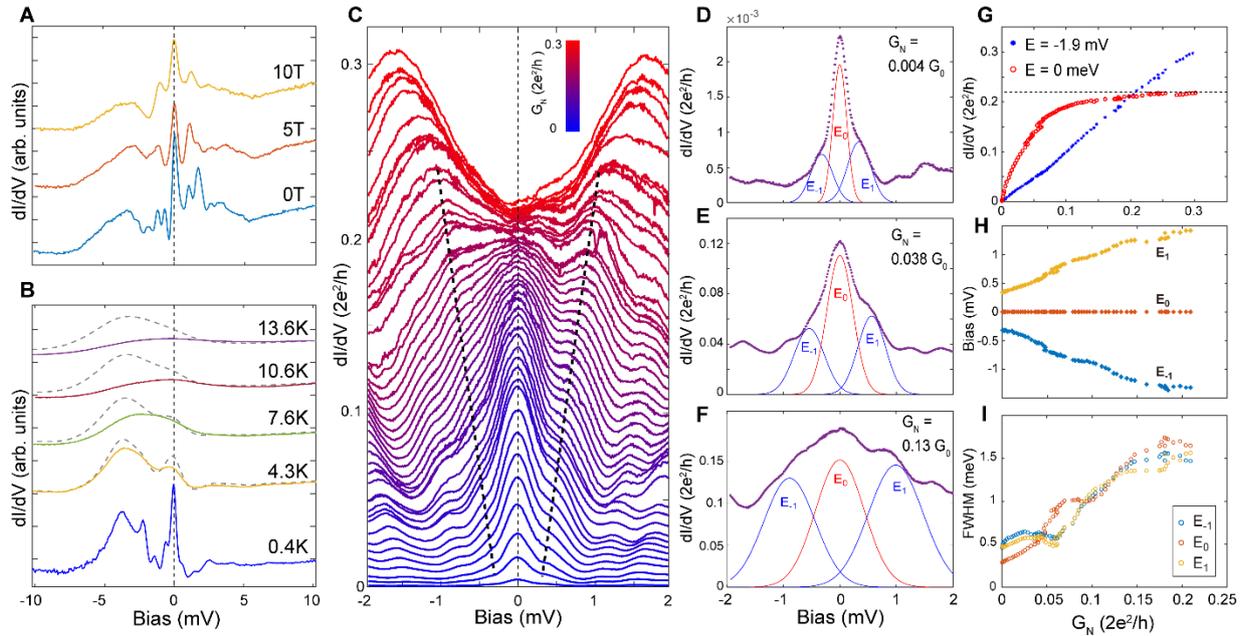

**Fig. 4. Magnetic field, temperature and tunneling transmission dependence of the in-gap states and ZECP.** (**A**) dI/dV spectra taken at the same quasiparticle puddle in 1UC MnTe under different magnetic field ($V_b$ =10 mV, $I$ =200 pA). (**B**) Temperature dependence of the dI/dV spectra with in-gap states. (Dashed curves are convolutions of the T=0.4K spectrum with the Fermi-Dirac distribution at elevated temperatures) (**C**) Evolution of dI/dV spectra with increasing tunneling transmission $G_N$ ($G_N$= $I_{set}/V_b$, $V_b$ = 2 mV). (**D-F**) Representative dI/dV spectra taken at different $G_N$, and the Gaussian fit to the three low-energy peaks (referred by $E_{\pm 1}$, $E_0$). (**G**) The tunneling conductance at 0 mV and -1.9 mV as a function of $G_N$. The absolute conductance value is calibrated by numerical differential of the I/V curve. (**H, I**) Energy positions and peak width (FWHM) of three fitted peaks ($E_0$, $E_{\pm 1}$) as a function of $G_N$, respectively.

It was shown that the tunnelling spectra at low barrier can help to clarify the nature of bound states in superconductors. *E.g.*, a Majorana mode should have quantized conductance of $2e^2/h$ at strong coupling region and sufficiently low temperature (45-47). We measured the dI/dV spectra at reduced tip-sample distance via gradually increasing $I_{set}$ at fixed $V_b$ of 2 meV ($G_N = I_{set}/V_b$ describes the tunneling transmission). The results are summarized in Fig. 4C. As $G_N$ increases, the ZECP maintains its position, while the side peaks gradually move to higher energy (tracked by dashed lines). The three representative spectra shown in Figs. 4D-4F that taken at different $G_N$ clearly display such behavior. Furthermore, when the conductance of ZECP reached about 0.22 ($2e^2/h$), it starts to saturate but the side peaks keep increasing and exceed the ZECP. In Fig. 4G we plot the ZBC as a function of $G_N$ and the saturation is directly evidenced. Similar behavior is also observed for other ZECPs (see Fig. S11 for another data set). We note such behavior resembles the vortex zero-modes reported in Fe(Te, Se), whose conductance are also saturated below $2e^2/h$ (48). It could be due to temperature broadening on a quantized peak, as the temperature here (T = 0.4K) is similar to the case in ref. 48.

To see detailed evolution of the in-gap states with increasing $G_N$, we performed Gaussian fit to the ZECP ($E_0$) and the two lowest side peaks (refer as $E_{\pm 1}$). The fitted peak positions and peak width (FWHM) as a function of $G_N$ are summarized in Figs. 4H and 4I, respectively. The FWHM of $E_0$ and $E_{\pm 1}$ states all increase with $G_N$ and they nearly have similar values, which indicates all these states have similar tunneling broadening at high $G_N$ (46). Thus the ZECP is distinguished from the $E_{\pm 1}$ peaks for its non-shifting behavior. We notice that topologically trivial YSR states were reported to have energy shift when the tip-sample distance changes (49), which was attributed to tip induced gating effect. In our heterostructure we also occasionally observed such YSR-like shifting peaks (shown in Fig. S12). Since the tunneling spectroscopy of 1UC MnTe (Figs. 1H and 2A,2B) evidenced a band gap and a shallow electron pocket with $k_F \approx 0.08$ (1/Å), which corresponds to a low carrier density, the tip gating could also be the cause of the shifting peaks like $E_{\pm 1}$. Therefore the non-shifting ZECP and its saturated conductance would suggest its nontrivial origin.

**Absence of in-gap states and gap enhancement in 2UC MnTe/Bi$_2$Te$_3$/Fe(Te, Se)**

We also measured the superconducting state in 2UC MnTe/Bi$_2$Te$_3$. Fig. 5A shows a topographic image which contains both 1UC and 2UC MnTe terraces. The height of second MnTe layer is similar to 1UC MnTe (Fig. 5C). Fig. 5B is the zero-bias dI/dV map taken on the surface of Fig. 5A. On the 1UC MnTe region, there are similar quasiparticle puddles in which ZECPs are found (Fig. 5D, points 2~4). However, for the 2UC MnTe region, no such in-gap states were observed. Its dI/dV spectra (Fig. 5E) display a uniform full superconducting gap with similar shape to that of 1UC Bi$_2$Te$_3$/Fe(Te, Se), which is also different from the enhanced gap observed in 1UC MnTe nearby (red curve in Fig. 5E, taken on point 1 in Fig. 5B). As the large scale spectrum of 2UC MnTe has a wide insulating gap with $E_F$ in the middle (Fig. 1H), the low-energy superconducting gap observed in 2UC MnTe is likely from direct tunneling into the underneath Bi$_2$Te$_3$/Fe(Te, Se). Thus these findings indicate that 2UC MnTe layer does not affect superconductivity of the system at all.

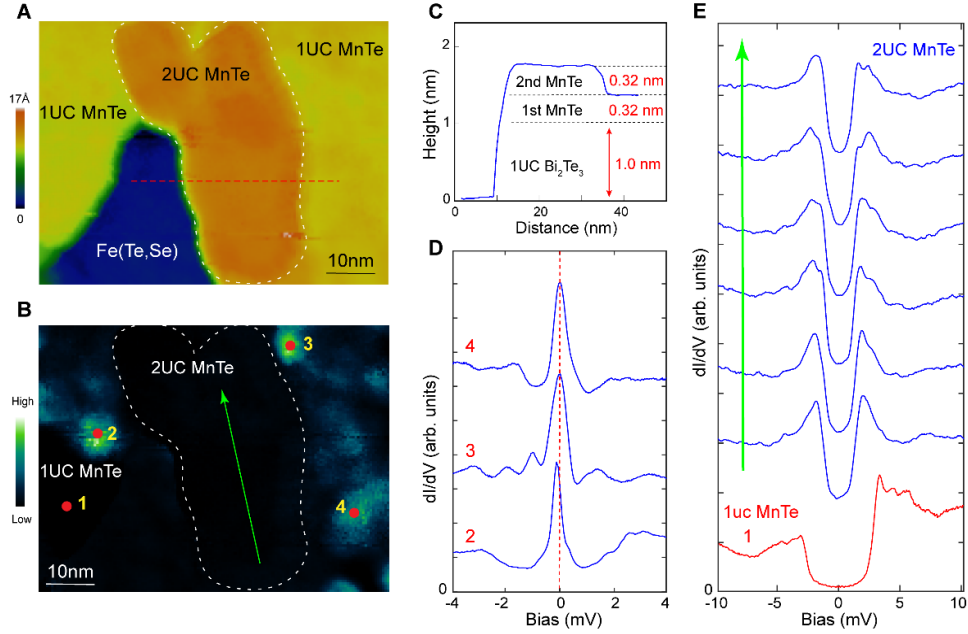

**Fig. 5. STM characterization of 2UC MnTe/Bi$_2$Te$_3$/Fe(Te, Se).** (**A**) Topological image including an 2UC MnTe terrace (53×70 nm$^2$, $V_b$=1 V, $I$=20 pA). The white dashed curve indicates the edge of 2UC MnTe terrace. (**B**) Zero bias dI/dV map taken in the area of panel (A) (setpoint: $V_b$=10 mV, $I$=150 pA). (**C**) The height profile along the red dashed line in panel (A). (**D**) dI/dV spectra measured at the 1UC MnTe regions which show a ZECP (on the spots 2-4 in panel B). (**E**) A series of dI/dV taken along the green arrow in panel B (on 2UC MnTe), and the spectrum taken on spot 1 in panel B (on 1UC MnTe). (setpionit: $V_b$=10 mV, $I$=150 pA)

**Calculations on the magnetism of 1UC and 2UC MnTe/Bi$_2$Te$_3$**

The quite different superconducting states in 1UC and 2UC MnTe imply that magnetism may play an important role. Since the bulk MnTe displays inter-layer AFM and intra-layer FM state (36,37), one may expect the magnetism in 1UC and 2UC MnTe is intrinsically different. However, it is also known that magnetism could change significantly in thin film or at interface. Therefore, we performed first-principle calculations on the magnetism of 1UC and 2UC MnTe/Bi$_2$Te$_3$ heterostructures (details are shown in Section S11). Firstly, sufficient structure optimization of 1UC and 2UC MnTe/Bi$_2$Te$_3$ were performed, and the results are shown in Figs. 6A-6C. They display similar in-plane lattice constant and inter-layer height as that measured by STM. Then we explore their most stable magnetic structures and evaluate the Heisenberg exchange coupling (J) in the LSDA+U framework. We find that 1UC MnTe/Bi$_2$Te$_3$ favors nearest-neighbor AFM coupling (AFM1 structure) and the magnetic moment of Mn lies in-plane; while the 2UC MnTe/Bi$_2$Te$_3$ favors an inter-plane AFM (AFM2) with out-of-plane orientated moments, as sketched in Figs. 6D-6E. The magnetic anisotropy energy of both the AFM1 and AFM2 states are larger than 5meV. Since MnTe has a triangular lattice, the nearest-neighbor AFM1 state will have intrinsic frustration.

The calculated AFM1 state in 1UC MnTe/Bi$_2$Te$_3$ is different from single UC MnTe in the bulk form. This is checked to be due to interfacial effect (e.g. lattice strain), as our calculation can correctly predict the in-plane AFM state of bulk MnTe (see SM). However, the above LSDA+U calculations only involve Heisenberg interactions, while for magnetic systems without inversion symmetry (such as heterostructures), SOC should give rise to DM interactions which favor twisted spin configuration. Based on a first-principles linear-response (FPLR) approach, we further calculated the DM interaction together with Heisenberg parameters of 1UC MnTe/Bi$_2$Te$_3$ (Section S11). The Heisenberg coupling (J) estimated by FPLR well agrees with that calculated by LSDA+U ($\approx$ 20.0 meV). Meanwhile, the strength of nearest-neighbor DM interaction $|D|$ is estimated to be 4.2 meV (specific values of the $D$ vector are shown in Table S6). Such a large DMI is originated from the large SOC of Bi$_2$Te$_3$ and the inversion asymmetric interface. We notice that several recent theoretical works on the structural asymmetric Mn dichalcogenide also give very large DMI (31,32).

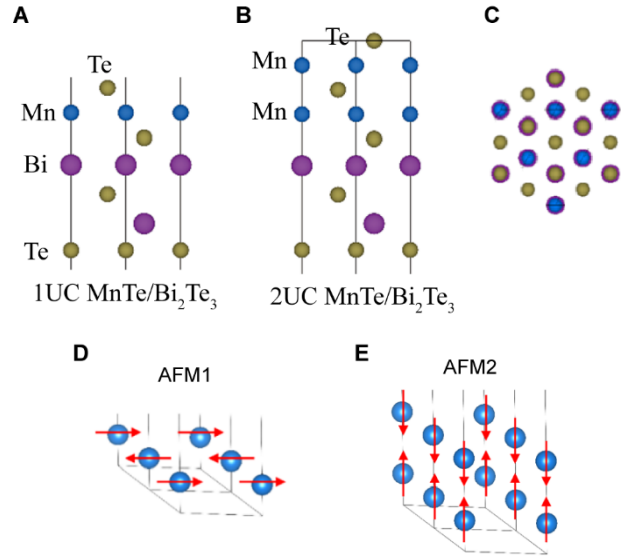

**Fig. 6. Calculated magnetic structure of 1UC and 2UC MnTe on Bi$_2$Te$_3$** (A, B) Side views of optimized structure of 1UC and 2UC MnTe/Bi$_2$Te$_3$ interface, respectively. The blue, purple and yellow balls represent Mn, Bi and Te atoms, respectively. (C) Top view of the interface. (D, E) Local magnetic structure of 1UC MnTe/Bi$_2$Te$_3$ and 2UC MnTe/Bi$_2$Te$_3$, respectively.

## Discussion

The robust ZECPs with non-trivial properties and enhanced superconducting gap in 1UC MnTe/Bi$_2$Te$_3$/Fe(Te, Se) are rather unconventional phenomena. Below we will discuss their origin. Firstly, the observed ZECPs are located inside of the MnTe terrace, not at the edge. This excludes their origin of chiral edge mode which was predicted for FM/TI/SC heterostructures (24,25). A natural explanation is that the 1UC MnTe here is not ferromagnetic as shown by above calculations. Secondly, such randomly distributed ZECPs seem to resemble the tunneling spectra of previously reported ferromagnet/superconductor heterostructures (5,6), in which ZECPs were also observed and were attributed to spin-triplet pairing state (1,2). However, we notice that such spin-triplet state would require a spin rotation at the interface (i.e., misalignment between the interfacial spin and bulk magnetization of FM layer), and therefore is usually sensitive to external magnetic field (5,6). In our measurement, the ZECP is robust against magnetic field (at least for out-of-plane field), and the MnTe layer is only 1UC thick, which is unlikely to give interfacial spin misalignment. Thus the spin-triplet pairing scenario is also not favored here.

Then the above calculations of magnetism can provide an alternative explanation. It was shown that DM interaction and frustrated AFM state can both promote (in-plane) non-collinear spin texture such as skyrmions (50-52). Here the calculated DMI strength in 1UC MnTe/Bi$_2$Te$_3$ is even larger than that of the Co/Pt ($\approx$ 3.0 meV) and Fe/Ir ($\approx$ 1.7 meV) thin films (53,54), in which skyrmions were observed with a typical size down to few nm. Recently, a number of theoretical works have shown that various types of non-collinear spin texture can induce unconventional pairings in magnet/superconducting interface, which often manifest themselves as in-gap states (10-19). Particularly, zero-energy modes can be induced inside of FM skyrmions (11,12,14) or at the end of AFM skyrmion chains under certain conditions (17). Here the ZECPs/in-gap states in 1UC MnTe/Bi$_2$Te$_3$/Fe(Te, Se) are distributed in separated regions with a typical size of 4~10 nm (Fig. 3 and Fig. S7). As defect induced YSR states are not the origin, they are likely induced by certain non-collinear spin textures residing in such length scale. Moreover, the in-plane orientated magnetic moments in AFM1 state with strong anisotropy energy (> 5 meV) suggest they should be robust against out-of-plane field, which is consistent with our measurement in Fig. 4A. For 2UC MnTe/Bi$_2$Te$_3$/Fe(Te, Se), the inter-plane AFM2 state with out-of-plane orientated moments is not frustrated and its interfacial DMI will also be weaker. Therefore, one expects to see less affected superconductivity in 2UC MnTe.

Besides, the enhanced superconducting gap in 1UC MnTe is another remarkable founding. To our knowledge, this is a very rare case that a proximity induced gap in magnetic material can be notably larger than the host superconductor. Most tunneling measurements on magnet/superconductor heterostructures have reported a reduced gap (5, 26-28). Here since a shallow electron pocket crossed $E_F$ in 1UC MnTe (Fig. 2B), which indicates the conducting electrons in 1UC MnTe participated in superconductivity. Thus our results indicate such interfacial 1UC MnTe layer is a "doped AFM insulator". We notice that the electron correlation in MnTe (in which the Mn atoms have 3d$^5$ configuration) was reported to be strong (55). The gap enhancement mechanism could be related to the frustrated magnetism and electron correlations in 1UC MnTe, which is quite interesting and worthy of further investigation. As for the 2UC MnTe, since its $E_F$ is well inside of the insulating gap and the magnetism is also different, the gap enhancement is absent.

At last, we shall point out another possibility for observing in-gap states at B= 0T: the "spontaneous" vortex state which was suggested for certain ferromagnet/superconductor interface (56) or superconductors with magnetic impurities (44,57,58). If assuming 1UC MnTe only provides an effective magnetic field to the underneath Bi$_2$Te$_3$/Fe(Te, Se) and generated the in-gap state/ZECP, one would expect to see similar ZECP in vortex of 1UC Bi$_2$Te$_3$ by applying an external field. We have checked the vortex states of 1UC Bi$_2$Te$_3$/Fe(Te, Se) (Fig. S6) which show that there is a broad peak near zero-bias at vortex center, and it gradually splits as leaving the vortex. This behavior is similar to that previously observed in Bi$_2$Te$_3$/Fe(Te, Se) (39,40) and Bi$_2$Te$_3$/NbSe$_2$ systems (59), but very different from the multiple in-gap states and non-splitting ZECP in 1UC MnTe here. Therefore the spontaneous vortex state scenario is also not favored.

In summary, we have successfully fabricated a tri-layer type magnet-superconductor heterostructure of MnTe/Bi$_2$Te$_3$/Fe(Te, Se) which has atomically sharp interface, and in which strong SOC is intentionally introduced. The observation of robust zero-energy modes and enhanced superconducting gap in 1UC MnTe region strongly suggest an unconventional pairing state. Via first-principle calculations, we show that 1UC MnTe/Bi$_2$Te$_3$ has large DMI due to its inversion asymmetric structure and strong SOC, which can generate non-collinear spin textures and give rise to ZECP. The enhanced gap could also be related to the frustrated magnetism and electron correlations in 1UC MnTe. Therefore, this heterostructure provides a promising platform for further exploring unconventional and topological nontrivial superconductivity.

**Materials and Methods:**

**Growth of the heterostructure:** The sample growth was conducted in a low-temperature STM system (UNISOKU 1300) equipped with a MBE chamber. The FeTe$_x$Se$_{1-x}$ (x ≈ 0.55) single crystal with T$_C$ = 14.5K was cleaved in UHV chamber at 78K. 1 UC Bi$_2$Te$_3$ was firstly grown on cleaved surface by co-depositing Bi (99.997%) and Te (99.999%) with a flux ratio of ≈ 1:10. Then 1~2 UC MnTe was grown by co-depositing Mn (99.95%) and Te (99.999%) with a flux ratio of ≈ 1:10. The substrate was kept at 200°C during the growth and was post-annealed at the same temperature for 10 min to further improve the quality. The sample was transferred *in-situ* from the MBE chamber to STM chamber after growth.

**STM measurement:** STM experiments were conducted at T= 0.4K or 4.2 K. PrIr tip was used after treatment on Au(111) surface. The dI/dV spectra were collected by a standard lock-in technique with a modulation frequency of 741 Hz and amplitude (ΔV) of 0.5 mV at 4.2 K and 0.1 mV at 0.4 K. The energy resolution and the zero point of sample bias were calibrated before measurement (Fig. S1 of SM).

**Acknowledgments:** We thank Prof. Xi Dai and T.K. NG for helpful discussions.

**Funding:** This work is supported by: National Key R&D Program of the MOST of China (Grant Nos. 2017YFA0303004, 2018YFA0704300); Innovation Program for Quantum Science and Technology (Grant No. 2021ZD0302800); National Natural Science Foundation of China (Grant Nos. 92065202, 11888101, 11790312, 11961160717, U1932217); Science Challenge Project (grant No. TZ2016004); Shanghai Municipal Science and Technology Major Project (Grant No. 2019SHZDZX01); Shanghai Pilot Program for Basic Research - Fudan University 21TQ1400100 (21TQ005); China Postdoctoral Science Foundation (Grant No. BX20200097).

**Author contributions:** The growth of heterostructure and STM measurement: S.Y.D., C.C., R.T., D.M.Z., Y.N.H., T.X.J., Y.J.Y. Fe(Te, Se) single crystal growth: Y.Q.P., Z.X.S. First-principle calculation: Z.P.C., D.W., X.G.W. Supervision: D.L.F, T.Z. Writing: D.L.F., X.G.W, T.Z. All authors have discussed the results and the interpretation.
**Competing interests:** The authors declare no competing interests.
**Data and materials availability:** All data needed to evaluate the conclusions in the paper are present in the paper and/or the Supplementary Materials.

\

# Supplementary Materials for
# Observation of robust zero-energy state and enhanced superconducting gap in a tri-layer heterostructure of MnTe/Bi₂Te₃/Fe(Te, Se)

**Section S1: Calibration of energy resolution at T=0.4K and bias offset.**

Limited by the thermal and electrical noise broadening, the energy resolution of STM system is estimated by $3.5k_B T_{eff}$ ($T_{eff}$ is the electronic temperature). To calibrate the $T_{eff}$, we measured the superconducting gap of Pb film grown on Si(111) substrate at 0.4K as shown in Fig. S1(A). BCS fitting gives $\Delta = 1.39$ meV, $T_{eff} = 1.18$ K and $\Gamma = 0.005$ meV and the energy resolution is $3.5k_B T_{eff} = 0.36$ meV.

Besides, the bias voltage ($V_b$) applied to the sample has a small offset which should be carefully calibrated to determine the zero-bias peak in our experiment. Therefore, we determine the true zero bias by measuring a set of $I$-$V$ curves taken at different setpoints, as they intersect at a single point where $V_b = 0$ and $I = 0$ (Fig. S1(B-C)). All the $dI/dV$ data in the paper are calibrated in the same way.

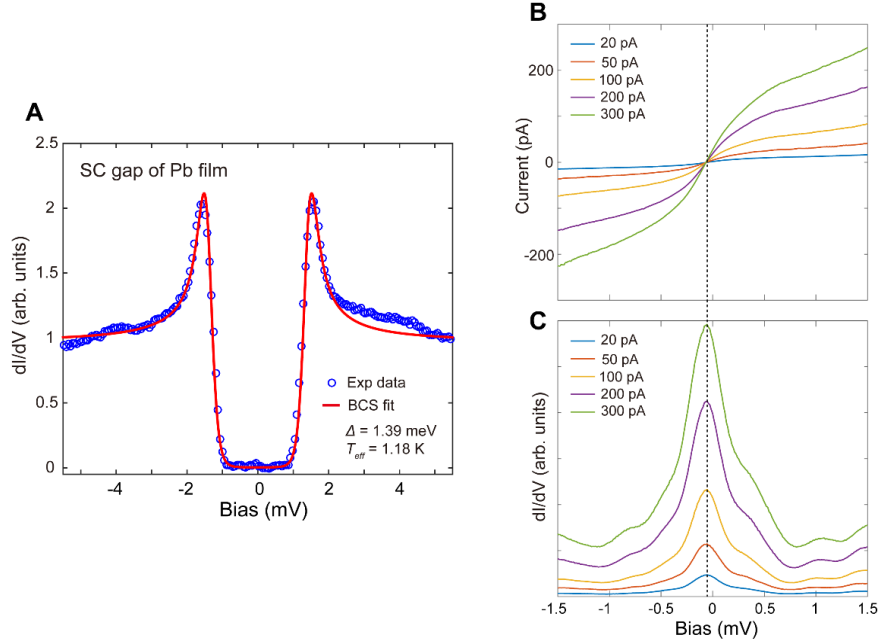

**Fig. S1. Calibration of energy resolution and bias offset.** (**A**) The superconducting gap of Pb film grown on Si(111). Blue circles: the experimental data measured at 0.4K. Red curve: BCS fit with $\Delta = 1.39$ meV, $T_{eff} = 1.18$ K and $\Gamma = 0.005$ meV. The energy resolution is $3.5k_B T_{eff} = 0.36$ meV. (**B**) A set of $I$-$V$ curves taken at different setpoints. The true zero bias is the intersection of these curves with offset -0.06mV. (**C**) The $dI/dV$ spectra obtained simultaneously with the I/V curves in panel (B).

**Section S2: Typical *dI/dV* spectra taken on surface defects.**

We measured the dI/dV spectra taken across the triangular shaped defects on MnTe and $Bi_2Te_3$ surfaces. As shown in Fig. S2(C-F), the spectra taken on these defect sites are very similar to that measured off defects. This means no additional impurity states are generated by these point defects. The peaks shown in Fig. S2(C-E) are actually the in-gap states from the quasiparticle puddle in 1UC MnTe, which shall originate from larger scale magnetic structure rather than single magnetic impurity.

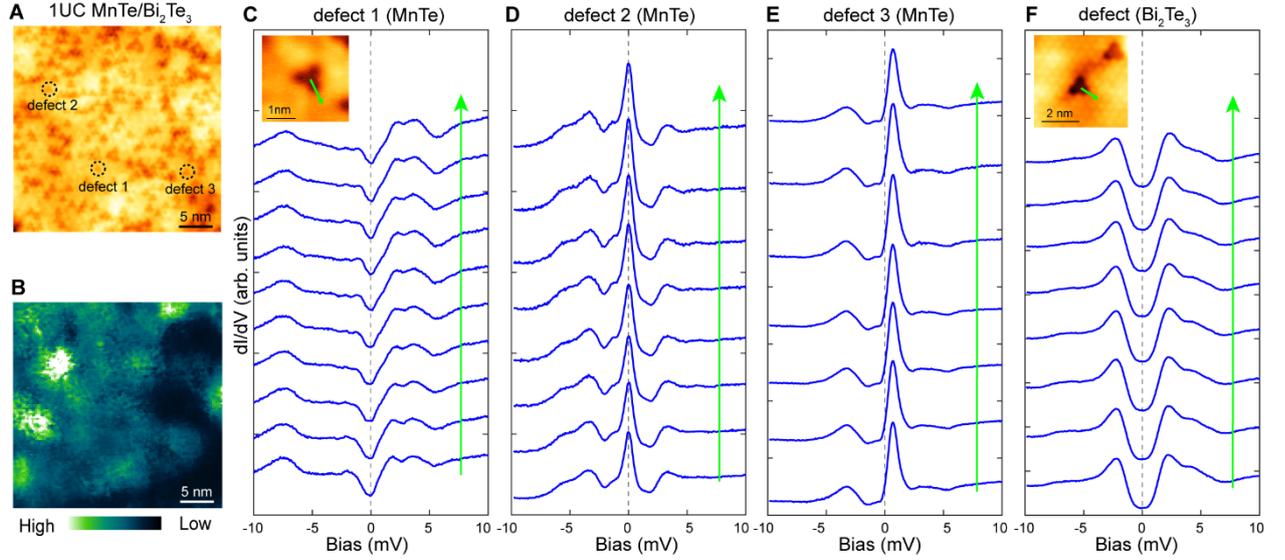

**Fig. S2. *dI/dV* spectra on the defects of 1UC MnTe and 1UC $Bi_2Te_3$ surface.** (A) STM topography of a 1UC MnTe surface ($33 \times 33 nm^2$, $V_b$ =1 V, $I$ =20 pA). (B) Zero-bias mapping taken at the same region in (A) ($V_b$=10 mV, $I$ =100 pA). (C-E) dI/dV spectra taken at three defects of 1UC MnTe surface indicated by the circles in (A) ($V_b$= 10mV, $I$ = 200pA). Inset: topography of defects 1 of MnTe surface. (F) *dI/dV* spectra measured at the defects on 1UC $Bi_2Te_3$ surface ($V_b$ = 10mV, $I$ = 200pA) with the corresponding defect image in the inset. All the data are taken at 0.4K and B = 0T.

**Section S3: Quasiparticle interference (QPI) measurement on 1UC MnTe and 1UC $Bi_2Te_3$.**

To examine the electronic state of 1UC MnTe and 1UC $Bi_2Te_3$, we performed dI/dV mappings and calculated their fast-Fourier transform (FFT) images, as shown in Fig. S3 and Fig. S4. For 1UC MnTe, six-fold scattering patterns can be seen in FFT. Via summarizing the FFT linecut across these scattering spots, an electron-like dispersion with band bottom around -60 meV and $q_F \sim 0.16$ (1/Å) is observed (Fig. S3B).

For 1UC $Bi_2Te_3$, the QPI patterns are also visible in dI/dV maps. A previous ARPES measurement (*Adv. Mater. 22, 4002 (2010)*) reported that 1UC 1UC $Bi_2Te_3$ still possesses surface state although it is gapped due to the hybridization. This is consistent with our QPI measurement here. A parabolic fitting to the dispersion extracted from FFT yields an electron-like band with a bottom at -250 (±10) meV and $q_F \sim 0.31$ (1/Å) (Fig. S4B). This dispersion is very different from that of 1UC MnTe.

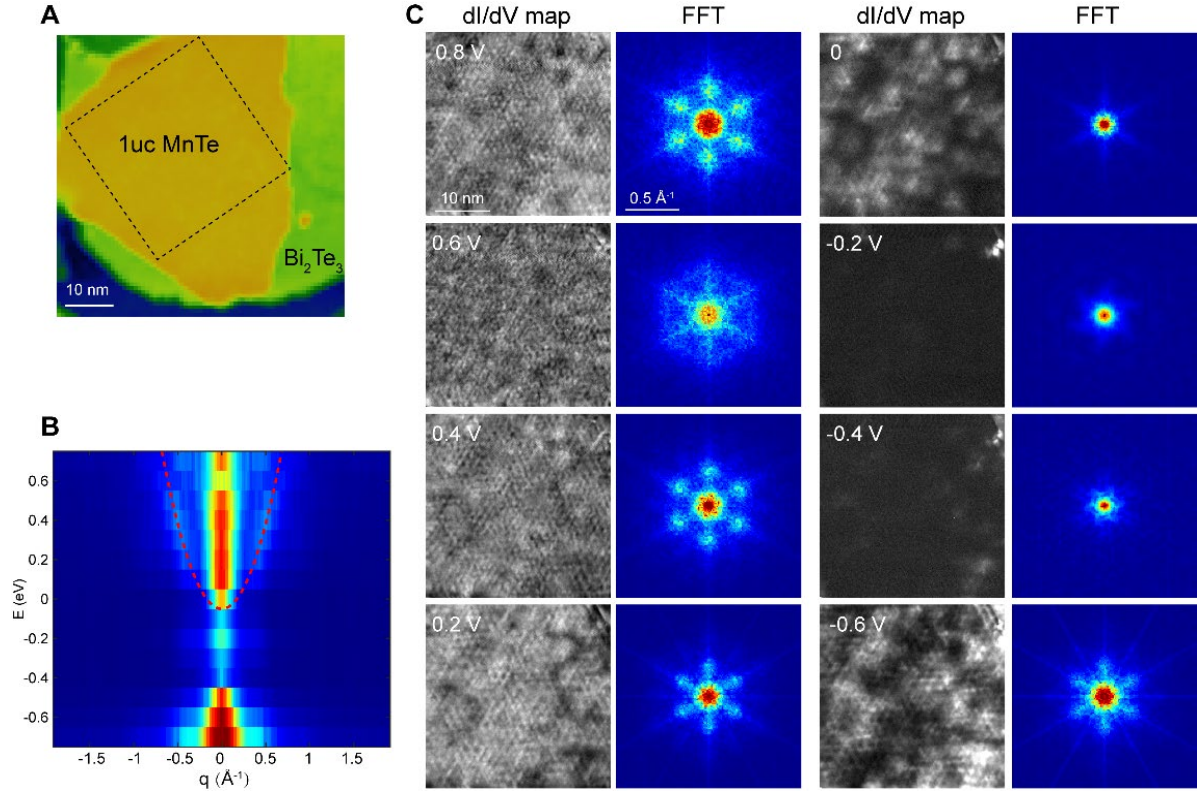

**Fig. S3. QPI measurement on 1UC MnTe.** (**A**) Topographic image of 1UC MnTe ($V_b = 1V$, $I = 20pA$). dI/dV maps are taken at the region of black dashed square (33×33 nm$^2$). (**B**) A summary of FFT line profile, in which an electron like dispersion can be seen with a band bottom at $E_b \approx -60$ meV and $q_F = 0.17$ (1/Å). (**C**) A series of dI/dV maps and corresponding FFT images taken at different bias (labeled on the map) ($I = 100pA$, 200×200 pixels). All the data are taken at 4.2K.

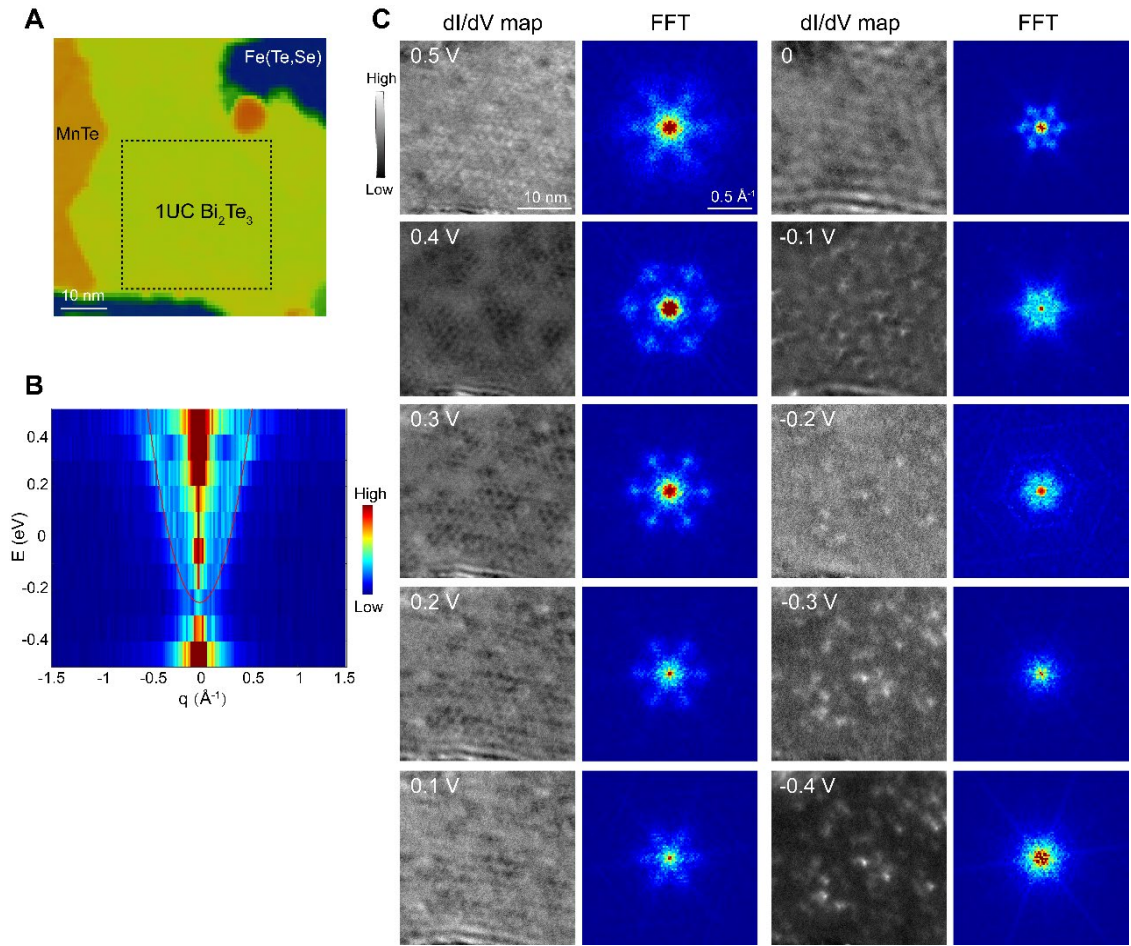

**Fig. S4. QPI measurement of 1UC Bi$_2$Te$_3$.** (**A**) Topographic image which shows a 1UC Bi$_2$Te$_3$ terrace ($V_b$ = 1V, I = 20pA). dI/dV maps in panel C are taken at the region in dashed square (33×33 nm$^2$). (**B**) A summary of FFT line profile, in which an electron like dispersion can be seen with a band bottom at $E_b \approx$ -250 meV and $q_F$ = 0.17 (1/Å). (**C**) A series of dI/dV maps and corresponding FFT images taken at different bias (labeled on the map). All the data are taken at 4.2K

**Section S4: Temperature dependence of the gap spectra taken on different surfaces.**

Fig. S5 shows the temperature dependent of gap spectra taken at different surface layers. The proximity superconducting gaps all disappear near the Tc (~14.5K) of Fe(Te, Se) substrate.

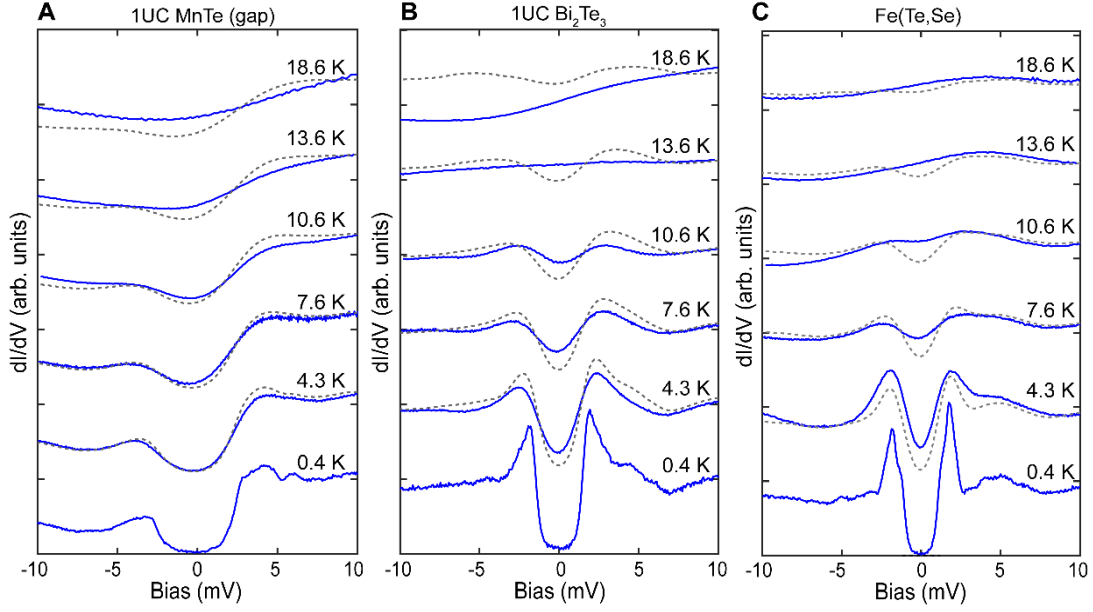

**Fig. S5. Temperature dependence of the gap spectra.** (A-C) The temperature dependence of superconducting gap on 1UC MnTe, 1UC Bi$_2$Te$_3$ and Fe(Te, Se), respectively (setpoint: $V_b$ = 10mV, $I$ =200 pA). Dashed curves in each panel are convolutions of the T=0.4K spectrum with the Fermi-Dirac distribution at elevated temperatures.

**Section S5: Vortex state measured on the gapped region of 1UC MnTe and 1UC Bi$_2$Te$_3$.**

We have observed the vortex state in some gapped region of 1UC MnTe, as shown in Fig. S6(A-C). Fig. S6A is a zero-bias dI/dV map taken at B=0 on a 1UC MnTe terrace, most of which has nearly zero ZBC. After applying B=10T out-of-plane field, a vortex core shows up in this region (indicated by dashed circle in Fig. S6B). Fig. S6C shows a series of dI/dV spectra taken across this vortex core. A pair of bound states are observed at the core center. They gradually disappear as leaving the center while the large gap of 1UC MnTe recovers. The emergence of vortex further confirms the superconducting nature of the large gap in 1UC MnTe.

Vortex cores can be observed on 1UC Bi$_2$Te$_3$ surface after applying 2T magnetic field as shown in Figs. S6(D-E). In Fig. S6F, a dI/dV linecut taken across the cortex exhibits a broad zero bias peak at the core center which split when leaving the center. These vortex states on Bi$_2$Te$_3$ surface are quite different from the discrete bound states on 1UC MnTe surface.

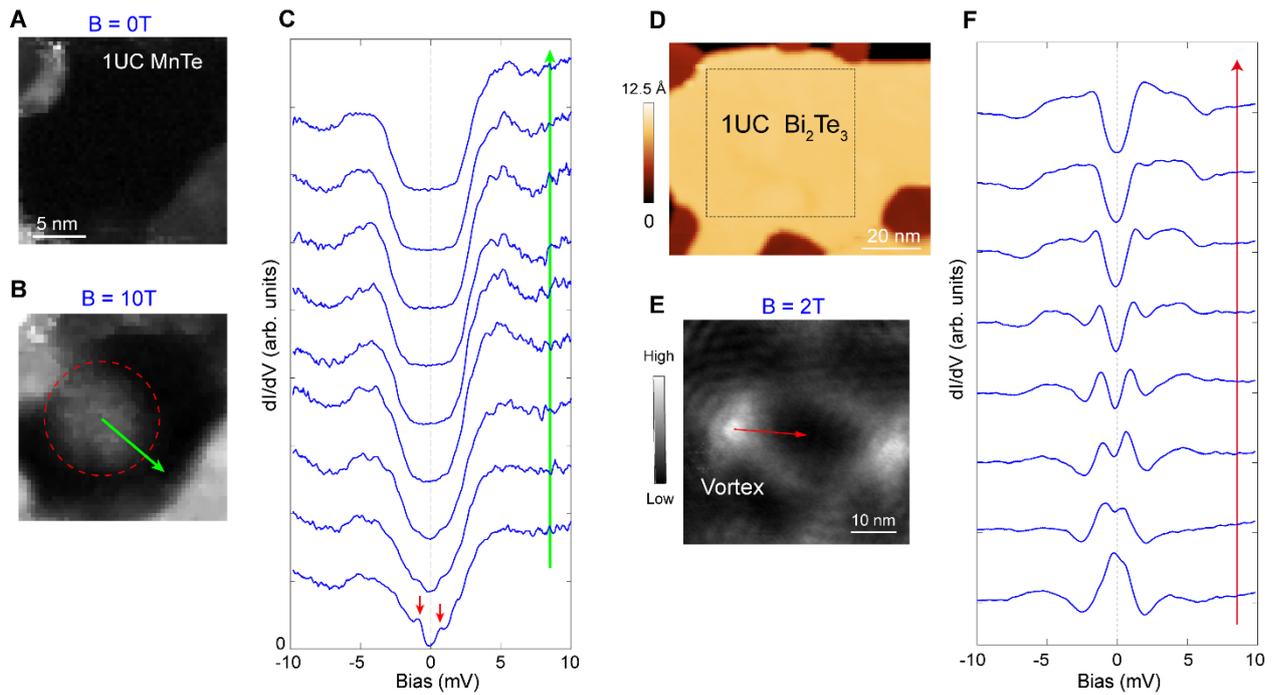

**Fig. S6. Vortex states measured on 1UC MnTe and 1UC Bi₂Te₃.** (**A, B**) Zero-bias dI/dV maps taken on 1UC MnTe surface (20 × 20 nm²) under B=0T and B = 10T, respectively. A vortex core shows up in panel B, marked by the dashed circle. (**C**) A series of dI/dV spectra taken along the green arrow in panel B (setpoint: $V_b$ = 10mV, I = 100pA). A pair of vortex state can be observed at the core center (red arrows). (**D**) Topographic image of 1UC Bi2Te3/Fe(Te,Se) ($V_b$ = 500 mV, I = 20 pA). (**E**) Zero bias dI/dV mapping taken at the region marked in panel (A) and B = 2T, which show two vortex cores ($V_b$ = 10 mV, I = 100pA). (**F**) dI/dV spectra taken along the arrow in panel (B) ($V_b$ = 10 mV, I = 200pA). All the data are taken at 0.4K.

**Section S6: Additional data set measured on different 1UC MnTe terrace.**

Fig. S7 shows another data set on 1UC MnTe terrace, quasiparticle puddles with multiple in-gap states and zero-bias peak are also observed.

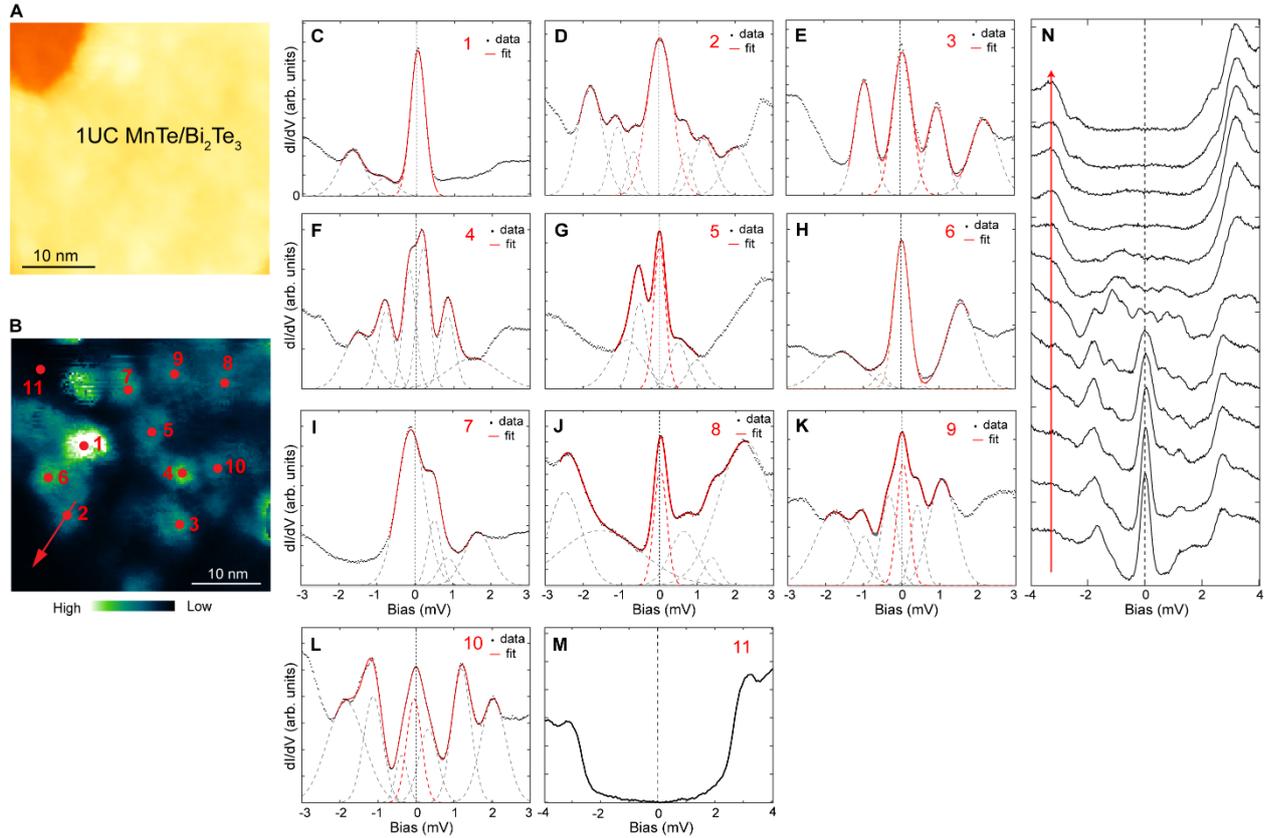

**Fig. S7. dI/dV spectra on 1UC MnTe surface.** (**A**) Topography of 1UC MnTe surface (40×40 nm$^2$, $V_b$ =1 V, $I$ =10 pA). (**B**) Zero bias mapping at the same region of panel (A). (**C-M**) *dI/dV* spectra taken at several positions marked in panel (B) ($V_b$=3 mV, $I$ =200 pA). (**N**) Linecut measure along the red arrow in panel (B) ($V_b$=4 mV, $I$ =200 pA). All the data are taken at 0.4K and B =0T.

## Section S7: Cross-correlation analysis of the surface defects (topography) and ZECP (zero-bias dI/dV map)

To quantitatively analyze the correlation between the surface defects and emergence of ZECP over large scale, we calculated the cross-correlation functions between the topographic image of 1UC MnTe and corresponding, as shown in Fig. S8 below. The correlation coefficient of +1 (-1) represents the perfect correlation (anti-correlation) between them, while a value approaching 0 indicates no correlation. Here we used two types of topography image to calculate cross-correlation: the first type was taken at Vb = 1V in which defects can be clearly resolved (Fig. S8A, D), and the second type (Fig. S8G) was taken at Vb = 10 mV in which defects are invisible but some surface unevenness can be seen. Such unevenness can also be induced by defects or was inherited from the underneath $Bi_2Te_3$/Fe(Te, Se). The zero-bias dI/dV maps of these surfaces are shown in Figs. S8(B, E, H). It is seen in Figs. S8(C, F, I) that for both kinds of these surface topography, the calculated correlation coefficients are small (< 0.12).

This indicates a rather weak correlation between surface defects/unevenness with the emergence of ZECP.

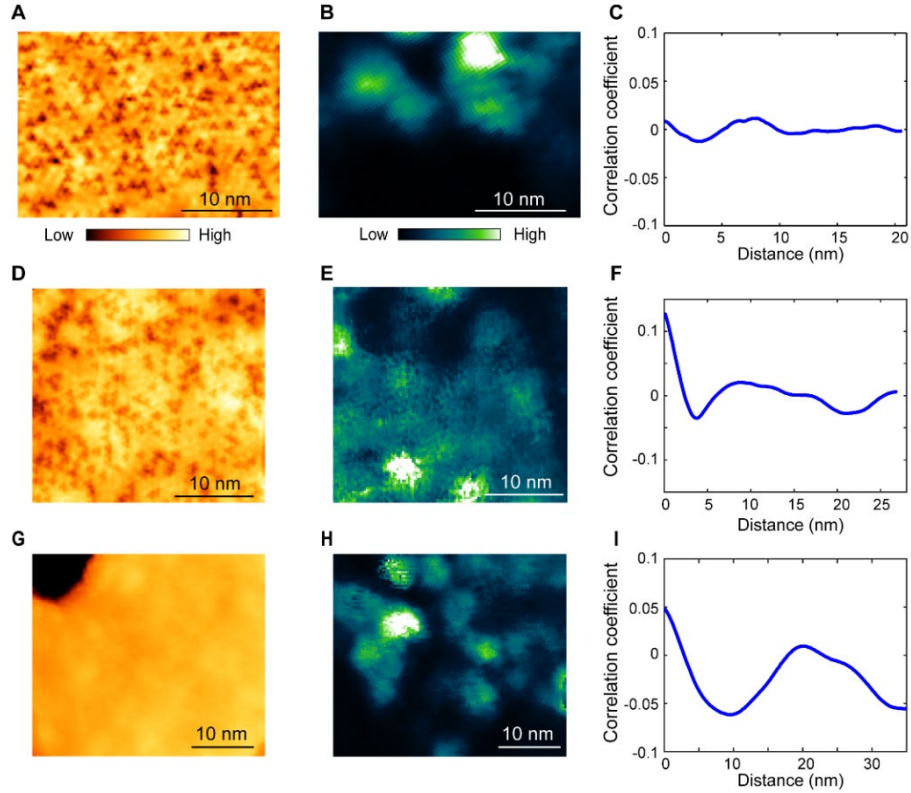

**Fig. S8. Cross-correlation analysis of the topography and zero bias map. (A, D)** Topographic images of MnTe surface taken at $V_b$ = 1V, in which triangular shaped point defect can be seen. **(G)** Topographic image of MnTe surface taken at $V_b$ = 10 mV. **(B, E, H)** Zero-bias dI/dV maps taken in the area of panel A, D, G, respectively. The quasiparticle "puddles" can be seen in all of them. **(C, F, I)** The azimuthally averaged cross-correlations between A and B, D and E, G and H, respectively.

**Section S8: Detailed analysis on the behaviors of zero/non-zero energy states under magnetic field, and the field dependence of the distribution of ZECP in 1UC MnTe.**

Comparing the behavior of non-zero and zero-energy states under magnetic field will be helpful to understand their nature. To do so we fitted the spectra of Fig. 4A (which clearly show multiple in-gap states) with multiple Gaussian peaks, as shown in Fig. S9. The fitted peak energies and width (FWHM) are summarized in Table. S1. It's seen that at B=0T (Fig. S9A), a ZECP and three pairs of non-zero states ($E_0$, $E_{\pm1}$, $E_{\pm2}$, $E_{\pm3}$) can be identified, which have nearly symmetric energies with respect to $E_F$. At high fields (Fig. S9B,C), the $E_0$ state can still be clearly seen, and its width (FWHM) only slightly increased from 0.55 meV for B=0 to 0.66 meV for B=5T, and 0.69 meV for B=10T (see Tab. S1). Such peak broadening (0.14 meV at 10T) is significantly smaller than the Zeeman energy (≈1.1meV at B=10T), indicative of non-splitting behavior. Meanwhile, the behaviors of non-zero states are very different from $E_0$. For instance, the $E_{\pm 1}$ peaks cannot be directly seen in the spectrum of B=5T, and only two broad non-zero peaks can be resolved at B=10T. By assuming the $E_{\pm 1}$, $E_{\pm 2}$, $E_{\pm 3}$ states still exist at high field, our fitting indicates the $E_{\pm 1}$ peaks shifted to higher energy at high field (Fig. S9B,C). The increase of the spacing between $E_{\pm 1}$ peaks ($E_{+1} - E_{-1}$) is 0.47 meV at B=5T and 1.21 meV at B=10T, which is close to the Zeeman energy. For the $E_{\pm 2}$ and $E_{\pm 3}$ states, their positions do not change much at high fields (Fig. S9B,C), but they are significantly broadened. The increasement of their FWHM are about 0.5-0.8 meV at B=10T, which could be an indication of peak splitting. Therefore, our analysis indicates the non-zero bound states are either shifted or significantly broadened by the magnetic field, while the ZECP are much less affected. This would again suggest a non-trivial origin of the ZECP.

The mappings on 1UC MnTe surface at different magnetic fields are shown in Figs. S10(A-D). The distributions of the quasi-particle puddles with high ZBC do not show obvious change. The magnetic dependence of dI/dV spectra taken on three different puddles (marked in Fig. S10B) are shown in Figs. S10(E-G).

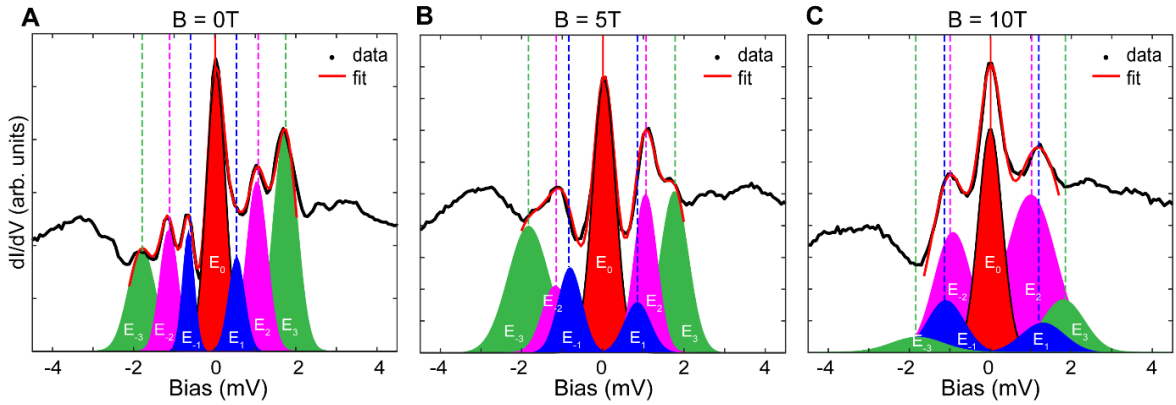

**Fig. S9. Multiple-Gaussian fitting of the spectra with different magnetic field. (A - C)**: dI/dV spectra taken at one quasiparticle puddle in 1UC MnTe under different magnetic field. The red curve are the multiple-Gaussian peaking fitting and individual peaks corresponding to $E_0$, $E_{\pm 1}$, $E_{\pm 2}$, $E_{\pm 3}$ are filled with different colors.

**Table S1.** Fitting parameters of the in-gap state measured under different magnetic field in Fig. S9. (Unit: meV)

|  |  | $E_0$ | $E_{+1}$ | $E_{-1}$ | $E_{+2}$ | $E_{-2}$ | $E_{+3}$ | $E_{-3}$ |
|---|---|---|---|---|---|---|---|---|
| B=0T | peak position | 0.03 | 0.54 | -0.65 | 1.04 | -1.13 | 1.74 | -1.80 |
| | FWHM | 0.55 | 0.48 | 0.35 | 0.58 | 0.51 | 0.67 | 0.77 |
| B=5T | peak position | 0.03 | 0.85 | -0.81 | 1.05 | -1.17 | 1.76 | -1.83 |
| | FWHM | 0.66 | 0.76 | 0.70 | 0.65 | 0.75 | 0.83 | 1.20 |
| B=10T | peak position | 0.00 | 1.30 | -1.10 | 1.02 | -0.92 | 1.80 | -1.80 |
| | FWHM | 0.69 | 1.25 | 1.08 | 1.40 | 1.08 | 1.21 | 1.50 |

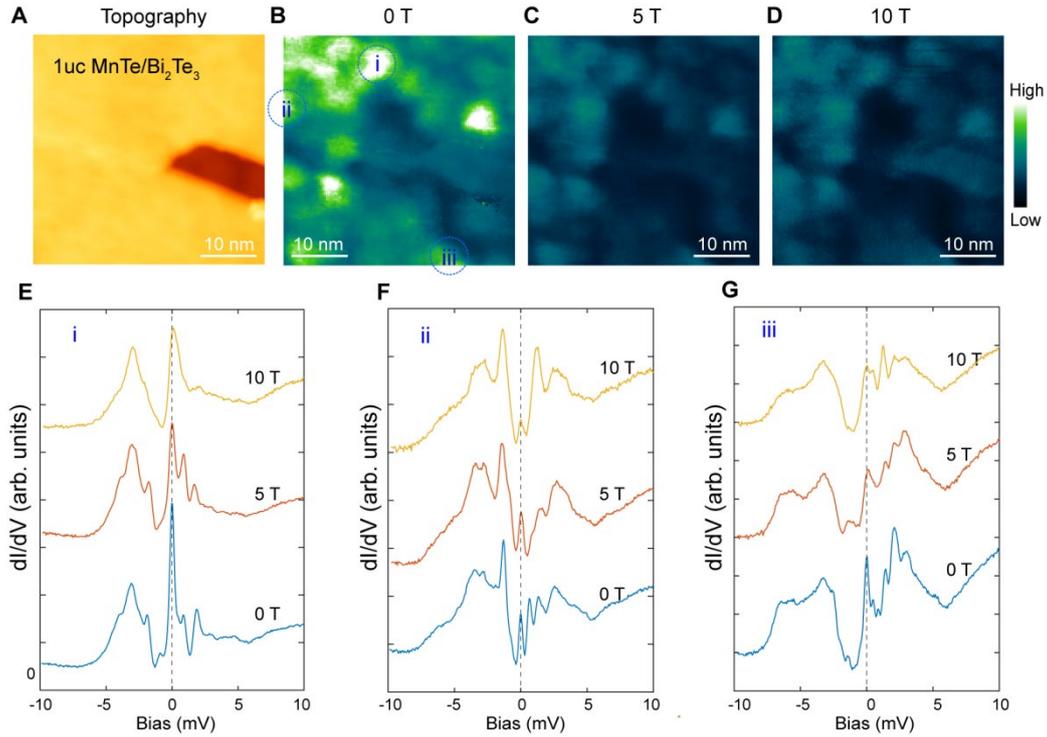

**Fig. S10. Magnetic dependence of the in-gap states/ZECP.** (**A**) The topography of 1UC MnTe surface (40 × 40 nm$^2$) ($V_b$ = 1V, $I$ = 20 pA). (**B-D**) Zero bias mapping at 0T, 5T and 10T magnetic field respectively. (**E-G**) $dI/dV$ spectra taken at three different locations marked in panel (B) ($V_b$ = 10 mV, $I$ = 200 pA, T=0.4K). The zero bias peak is robust without splitting even at 10T.

**Section S9: Additional data of the evolution of the ZECP at reduced tunneling barrier.**

In Fig. S11A below we plot another set of dI/dV spectra taken at reduced tip-sample distance (increased tunneling transmission $G_N$). It displays similar behavior of that shown in Fig. 4(C). The two non-zero energy side peaks gradually shift to higher energy when $G_N$ increases, while the ZECP stays at zero (peak positions are summarized in Fig. S11D). The tunneling conductance at E=0 also shows a trend of saturation when $G_N$ > 0.2 (2e$^2$/h), but the conductance of side peaks keep increasing (Fig. S11C).

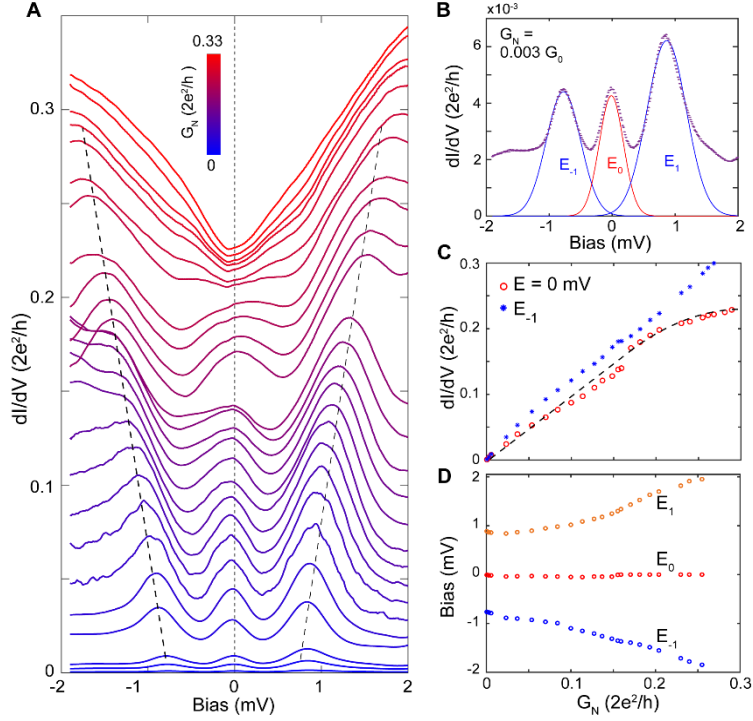

**Fig. S11 Tunneling transmission dependence of the in-gap states and zero energy mode.** (**A**) Evolution of dI/dV spectra with increased tunneling transmission $G_N$ ($G_N = I_{set}/V_b$), measured in another quasiparticle "puddle" with a ZECP and two side peaks. (**B**) A representative dI/dV spectra taken at low $G_N$, and the Gaussian fit to the three low-energy peaks. (**C**) The tunneling conductance at E=0 and of the $E_{-1}$ peak, as a function of $G_N$. (**D**) Energy positions of the in-gap peaks ($E_0$, $E_{\pm 1}$), as a function of $G_N$.

**Section S10: Impurity induced trivial in-gap states (coexisting with the ZECP).**

We occasionally observed some ring-like structure in zero-bias mapping of 1UC MnTe region, as that shown in Fig. S12(A) (Topographic image of this region is shown in Fig. S12(B)). Fig. S12(C) displays a series of *dI/dV* spectra taken across the "ring" (along the arrow in Fig. S12(A)). A peak whose energy keeps shifting with the change of tip position is observed. A color plot of Fig. S12(C) is shown in Fig. S12(D), and one can clearly see that the shifting peak went across $E_F$ at certain tip position, which results the ring in the zero-bias DOS mapping. Moreover, we found this peak also shifts notably when the vertical tip-surface distance changes (Fig. S12(E)). This indicates a tip induced local gating effect. We noticed ref. 49 has reported similar "spatially dispersive" state in pristine Fe(Te,Se) which was attributed to a YSR state caused by an impurity. Thus, it is very likely that such a shifting peak is induced by some underneath defects in the heterostructure (which is not evidenced in STM image). Meanwhile, one can still see a zero-bias peak in the spectra in Fig. S12(C-E), and in particular, it does not shift (or split) when the tip position or tip-surface distance changes. This again suggest the nontrivial origin of the zero-bias peak.

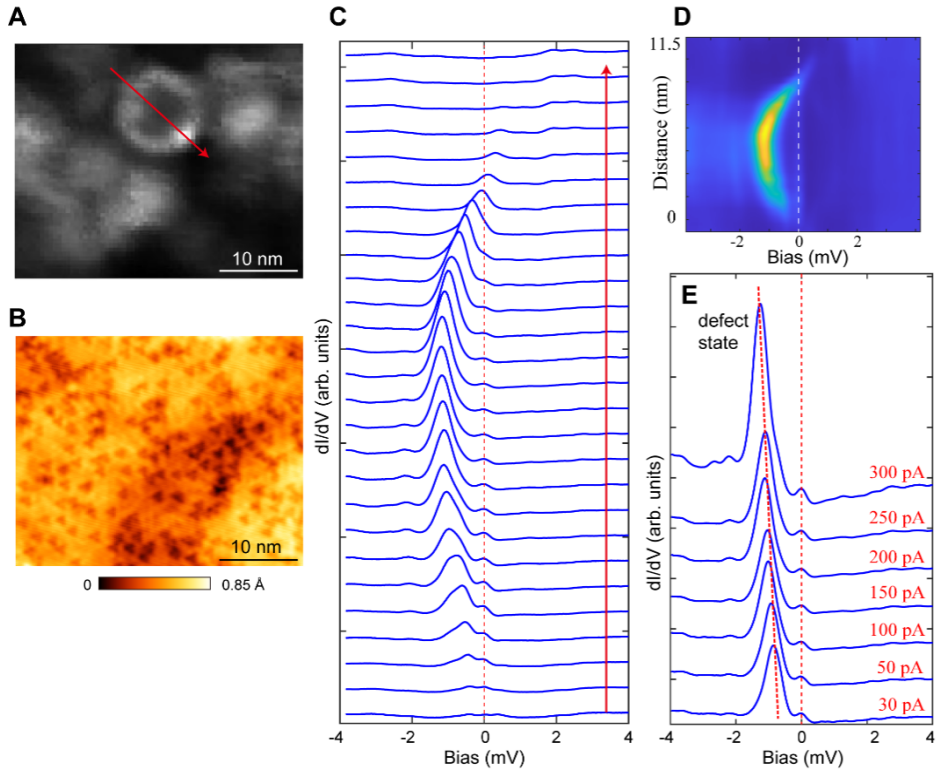

**Fig. S12. Impurity induced trivial in-gap states.** (**A**) Zero-bias dI/dV mapping taken on a 1UC MnTe region which displays a ring like structure ($V_b$ =10 mV, $I$ =100 pA). (**B**): Topographic image of the region in (A) ($V_b$ = 1V, $I$ = 30 pA). (**C**): A set of *dI/dV* spectra taken along the arrow in panel (A). (**D**): color plot of panel (C). (**E**): A series of *dI/dV* spectra taken at the same position inside of the "ring" in (A), but at different tunneling currents (different tip-surface distance).

**Section S11: Calculations on the magnetism of 1UC and 2UC MnTe/Bi$_2$Te$_3$**

Before studying the magnetism of heterostructures, we first check our calculation to predict the magnetism of bulk MnTe. Here four magnetic configurations as shown in Fig. S13 are taken into account, and their energies are determined by LSDA+U calculations. The relative total energies of unit cell are 0.107, 0, 0.081, 0.026 eV for FM, AFM1, AFM2, AFM3 configurations in Fig. S13, respectively. We find that the AFM1 configuration has the lowest energy which is consistent with the experiment (ref. 36, 37).

To explore the magnetism in the heterostructure, we performed the first-principles calculations for 1UC and 2UC MnTe/Bi$_2$Te$_3$. First of all, considering that the atoms on the plane perpendicular to **c** axis form the hexagonal lattice for both bulk MnTe and Bi$_2$Te$_3$, there is a chance to form stable interface by means of these two materials. Furthermore, the hexagonal lattice constants of them are similar, where the constant of MnTe is $a_M$=4.193Å and that of $Bi_2Te_3$ is $a_B$=4.369Å. The corresponding mismatch rate is 2|$a_M$-$a_B$|/($a_M$+$a_B$)×100%=4.11%. The mismatch rate smaller than 5% means that stacking the hexagonal unit cells of MnTe and $Bi_2Te_3$ alone **c** axis to form the heterostructure is reasonable. The lattice constant of heterostructure before the structure optimization is set to the average of $a_M$ and $a_B$. Then, due to the different end faces of these two materials, we consider six possible scenarios for 1UC and 2UC heterostructures, respectively (Fig. S14). To determine the most stable structure, we perform sufficient structural optimizations for these possible heterostructures. According to the total energy after optimization, we find that the M2B2 structures are more stable for both 1UC and 2UC cases. Therefore, the optimized M2B2 structures are used for following magnetic calculations.

Then the magnetism of optimized M2B2 structures is studied. We first consider different magnetic configurations of Mn atoms in the cases of 1UC and 2UC (Fig. S15), and perform corresponding LSDA+U calculations. The total energy and magnetic moments on Mn atoms are demonstrated in Tables S2 and S3 for 1UC and 2UC, respectively. We find that for different U values, the configurations AFM1 and AFM2 are more stable for 1UC and 2UC, respectively. For the magnetic moment of Mn, its magnitude increases when U becomes larger. In addition, in the case of 2UC, the magnetic moments of Mn atoms belong to different planes are different. To be specific, the Mn close to $Bi_2Te_3$ layer has a larger magnetic moment.

By mapping these total energies for different magnetic configurations obtained with LSDA+U calculations in Tables S2 and S3, the exchange couplings could be estimated by the energy-mapping analysis (*H. Xiang, et al., Dalton Transactions 42, 823 (2013)*). These magnetic systems are described by the Heisenberg Hamiltonian $\sum_{i<j} J_{ij} e_i \cdot e_j$, where $J_{ij}$ stands for the spin exchange parameter, while $e_i$ and $e_j$ represent the unit vectors in the direction of local magnetic moments on the site $i$ and $j$. For 1UC heterostructure, we only consider the nearest-neighbor interactions in ab-plane (labeled as J1), while for 2UC heterostructure, we consider the in-plane and interplanar nearest-neighbor interactions, which are labeled as J1 and J2, respectively. The calculated exchange couplings from the energy mapping analysis are summarized in Table S4. Note that for 1UC heterostructure, the nearest-neighbor Mn atoms in ab-plane favor antiferromagnetically. On the contrary, the nearest-neighbor interactions in ab-plane for 2UC heterostructure are ferromagnetic. This may be due to the structural difference for 1UC and 2UC heterostructure. Meanwhile, as shown in Table S4, the interplanar interactions along **c** axis are antiferromagnetic, which makes that AFM2 has the lowest energy in these three magnetic configurations for 2UC heterostructure.

The anisotropy of relatively stable magnetic configurations for 1UC and 2UC (AFM1 and AFM2) is further studied. Here four directions of moments as shown in Fig. S16 are taken into account, and the corresponding total energy is also calculated. We set U = 5eV and the spin-orbit coupling effect is also considered. According to the results of total energy listed in Table S5, we find that the moment along (100) has the lowest energy for 1UC. However, it becomes the (001) direction for 2UC. This phenomenon implies the change of easy magnetization axis with increasing the layer of MnTe. Furthermore, the magnitude of Mn almost unchanged under different moment directions for both 1UC and 2UC, which indicates the rationality of determining the magnetic anisotropy through comparing total energy. We find that the Mn close to $Bi_2Te_3$ layer also has a larger magnetic moment in 2UC, which is the same as our results of LSDA+U calculations.

Due to the lack of inversion symmetry in these magnetic systems, spin-orbit coupling can give rise to the Dzyaloshinskii-Moriya (DM) interactions (ref. 30). The DM interaction favors twisted spin structures and always plays a crucial role in many classes of magnetic systems such as skyrmion formation. Using the WIEN2K package and based on a first-principles linear-response (FPLR) approach (*X. Wan et al., PRL 97, 266403 (2006)*), we calculate the Heisenberg and DM interaction parameters in 1UC heterostructure with LSDA+U (U=5eV) calculations. The nearest-neighbor calculated exchange couplings from FPLR approach are summarized in Table S6. The Heisenberg interaction parameter is estimated to be 20.0 meV, which is well agreement with the calculated value 19.9 meV by the energy-mapping analysis with LSDA+U (U=5eV) calculation (see Table S4). Meanwhile, the calculated nearest-neighbor DM interaction parameters for different directions are listed in Table S6. It can be seen that the strength of DM interaction $|D|$ is about 4.2 meV. Thus the ratio $|D|/J$ is relatively large (about 0.21), which makes this material to be a candidate for skyrmion with small size.

The first-principles calculations are based on the density functional theory (DFT) implemented by Vienna Ab initio Simulation Package (VASP) (*Kresse, Comput. Mater. Sci. 6 15 (1996)*). The Perdew-Burke-Ernzerhof (PBE) functional of generalized gradient approximation (GGA) is chosen as the exchange-correlation potential (*Perdew et al., PRL 77 3865(1996)*), and the projector augmented wave (PAW) method is used to treat core-valence electron interactions (*Blochl, PRB 50 17953 (1994)*). The plane wave cutoff energy is set to 410 eV. For LSDA+U calculations, the rotationally invariant approach is adopted (*Dudarev et al., PRB 57 1505 (1998)*). In the magnetic calculations of bulk MnTe, the U is set to 5 eV. The unit cell of bulk MnTe combined with a 21×21×10 *k* mesh are used for FM and AFM1 configurations, and the 1×√3×1 supercell combined with a 16×9×10 *k* mesh are used for AFM2 and AFM3 configurations. In the calculations of heterostructures, the *k* mesh is set to 18×18×1. The vacuum layer is 20 Å for each heterostructure. Structure optimization is completed by the conjugate gradient method, where the **c** axis of cell is fixed and the convergence of force on each atom is less than 0.01 eV/Å. In the calculations of AFM1 and AFM3 configurations for heterostructures, the 1×2×1 supercell is used, and the corresponding mesh of Brillouin zone is changed to 18×9×1.

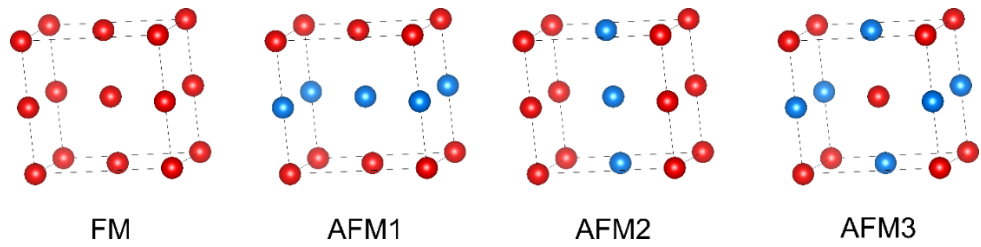

**Fig. S13. Four magnetic configurations of Mn atoms in bulk MnTe.** The red and blue colors represent spin up and down, respectively.

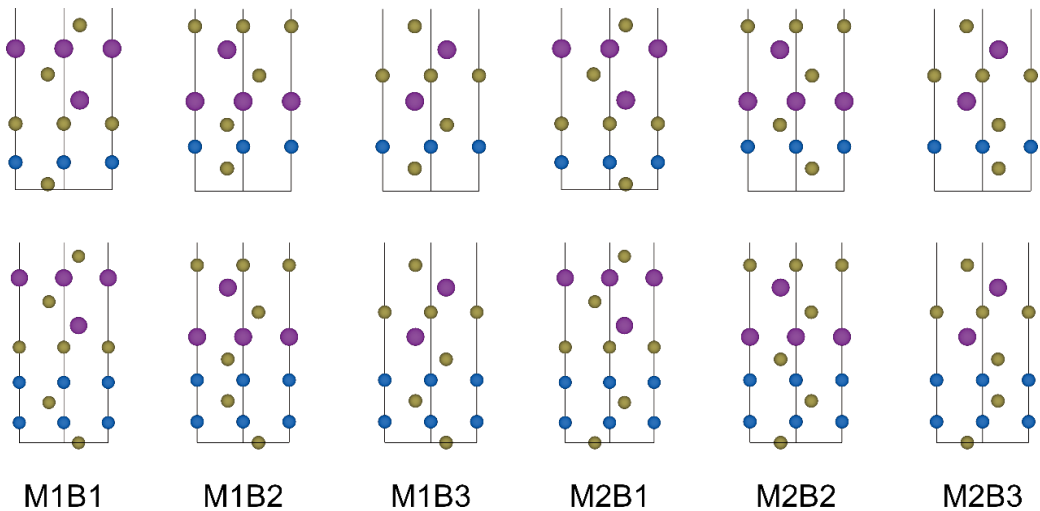

**Fig. S14. The side view along [110] directions for possible structures of 1UC and 2UC MnTe/Bi$_2$Te$_3$ heterostructures.** Each column represents a type of stacking MnTe and Bi$_2$Te$_3$ layers, which is named in the form of "MxBy". The first row represents the case of 1UC and the second row corresponds to 2UC. The blue, purple and yellow balls represent Mn, Bi and Te atoms, respectively.

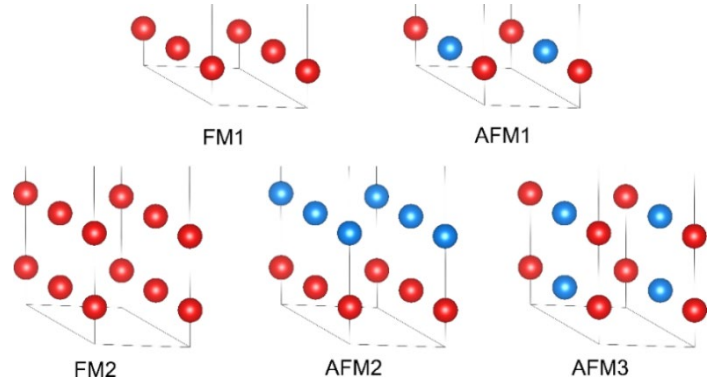

**Fig. S15. Different magnetic configurations of Mn atoms in 1UC and 2UC heterostructures.** The first row represents the configurations of 1UC and the second row corresponds to 2UC. The red and blue colors represent spin up and down, respectively.

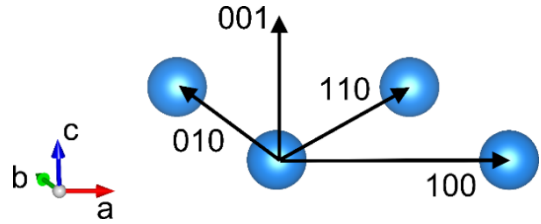

**Fig. S16. Different directions of magnetic moments on Mn atoms.**

| U (eV) | AFM1 (eV) | FM1 (eV) | AFM1 (μB) | FM1 (μB) |
|---|---|---|---|---|
| 4 | 0 | 0.102 | 4.14 | 4.22 |
| 5 | 0 | 0.080 | 4.25 | 4.31 |
| 6 | 0 | 0.062 | 4.35 | 4.40 |
| 7 | 0 | 0.049 | 4.43 | 4.47 |
| 8 | 0 | 0.039 | 4.50 | 4.53 |

**Table S2.** The relative total energy of unit cell and magnetic moment of Mn for 1UC heterostructure.

| U (eV) | FM2 (eV) | AFM2 (eV) | AFM3 (eV) | FM2[a] (μB) | | AFM2 (μB) | | AFM3 (μB) | |
|---|---|---|---|---|---|---|---|---|---|
| 4 | 0.126 | 0 | 0.235 | 3.78 | 3.84 | 3.94 | -4.01 | 3.91 | 4.02 |
| 5 | 0.169 | 0 | 0.224 | 4.10 | 4.17 | 4.09 | -4.15 | 4.10 | 4.19 |
| 6 | 0.148 | 0 | 0.203 | 4.27 | 4.33 | 4.21 | -4.27 | 4.23 | 4.31 |
| 7 | 0.118 | 0 | 0.181 | 4.37 | 4.42 | 4.32 | -4.37 | 4.33 | 4.40 |
| 8 | 0.092 | 0 | 0.161 | 4.45 | 4.50 | 4.40 | -4.45 | 4.42 | 4.48 |

**Table S3.** The relative total energy of unit cell and magnetic moment of Mn for 2UC heterostructure. [a]The two columns of data represent the magnetic moments of Mn atoms belong to different planes perpendicular to **c**, and the second one more closer to the $Bi_2Te_3$ layer. This convention is the same for AFM2 and AFM3.

| | 1UC | 2UC | |
|---|---|---|---|
| U (eV) | J1 (meV) | J1 (meV) | J2 (meV) |
| 4 | 25.6 | -13.7 | 62.9 |
| 5 | 19.9 | -6.93 | 84.5 |
| 6 | 15.6 | -6.79 | 74.2 |
| 7 | 12.3 | -7.85 | 59.1 |
| 8 | 9.66 | -8.62 | 45.9 |

**Table S4.** The calculated exchange couplings by the energy-mapping analysis for 1UC and 2UC heterostructure.

| Directions | 001 | | 100 | | 010 | | 110 | |
|---|---|---|---|---|---|---|---|---|
| 1UC (meV) | 5.231 | | 0 | | 6.546 | | 6.531 | |
| 1UC (μB) | 4.25 | | 4.24 | | 4.25 | | 4.25 | |
| 2UC (meV) | 0 | | 8.659 | | 8.657 | | 8.657 | |
| 2UC (μB)[a] | 4.08 | -4.16 | 4.08 | -4.16 | 4.08 | 4.16 | 4.08 | 4.16 |

**Table S5.** The relative total energy and magnetic moment of Mn for different directions of AFM1(1UC) and AFM2(2UC) magnetic configurations under U = 5eV. ([a]The two columns of data for one direction represent the magnetic moments of Mn atoms belong to different planes perpendicular to **c**, and the second one closer to the $Bi_2Te_3$ layer.)

|  | $\vec{d}$ | calculated value(meV) |
|---|---|---|
| J |  | 20.0 |
| DM | $\left(\frac{\sqrt{3}}{2}, -\frac{1}{2}, 0\right)$ | (0.6, 1.7, 3.5) |
|  | $\left(\frac{\sqrt{3}}{2}, \frac{1}{2}, 0\right)$ | (-0.6, 1.7, -3.5) |
|  | (0,1,0) | (-2.3, 0, 3.5) |
|  |  | $|D|$ = 4.2 meV |

**Table S6.** The nearest-neighbor calculated exchange couplings from FPLR approach for 1UC heterostructure with LSDA+U (U=5eV) calculation. The unit of distance $\vec{d}$ is taken as the lattice constant $a$.